\documentclass[10pt,aps,prb,twocolumn,nofootinbib]{revtex4}

\usepackage{amssymb,amsmath}
\usepackage{subfig}
\usepackage{amsmath}
\usepackage{graphicx}
\usepackage{color}
\usepackage{ragged2e}
\usepackage{hyperref}
\usepackage{etoolbox}
\makeatletter
\patchcmd{\NAT@citesuper}
      {\textsuperscript{#1}}{[\color{blue}{#1}\color{black}]}{}{}
\usepackage{listings}

\hypersetup{
    bookmarks=false,         
    unicode=false,           
    pdftoolbar=false,        
    pdfmenubar=false,        
    pdffitwindow=false,      
    pdfstartview={FitH},     
    colorlinks=true,         
    citecolor=blue,         
    urlcolor=blue
}

\begin{document}
\title{Spin injection and pumping generated by a direct current\\ 
flowing through a magnetic tunnel junction}
\author{A. I. Nikitchenko$^{1, 2}$ and N. A. Pertsev$^1$}
\affiliation{$^1$Ioffe Institute, St. Petersburg 194021, Russia\\
$^2$Peter the Great St. Petersburg Polytechnic University, St. Petersburg 195251, Russia
}

\begin{abstract}
A charge flow through a magnetic tunnel junction (MTJ) leads to the generation of a spin-polarized current which exerts a spin-transfer torque (STT) on the magnetization. When the density of applied direct current exceeds some critical value, the STT excites high-frequency magnetization precession in the "free" electrode of MTJ. Such precession gives rise to microwave output voltage and, furthermore, can be employed for spin pumping into adjacent normal metal or semiconductor. Here we describe theoretically the spin dynamics and charge transport in the CoFeB/MgO/CoFeB/Au tunneling heterostructure connected to a constant-current source. The magnetization dynamics in the free CoFeB layer with weak perpendicular anisotropy is calculated by numerical integration of the Landau-Lifshitz-Gilbert-Slonczewski equation accounting for both STT and voltage controlled magnetic anisotropy associated with the CoFeB$|$MgO interface. It is shown that a large-angle magnetization precession, resulting from electrically induced dynamic spin reorientation transition, can be generated in a certain range of relatively low current densities. An oscillating spin current, which is pumped into the Au overlayer owing to such precession, is then evaluated together with the injected spin current. Considering both the driving spin-polarized charge current and the pumped spin current, we also describe the charge transport in the CoFeB/Au bilayer with the account of anomalous and inverse spin Hall effects. An electric potential difference between the lateral sides of the CoFeB/Au bilayer is calculated as a function of distance from the CoFeB$|$MgO interface. It is found that this transverse voltage signal in Au is large enough for experimental detection, which indicates significant efficiency of the proposed current-driven spin injector. 
\end{abstract}

\maketitle
\section{\label{sec:one}Introduction}

In conductive ferromagnetic nanolayers, magnetic dynamics can be induced by a spin-polarized charge current exerting a spin-transfer torque (STT) on the magnetization ~\cite{Slonczewski1989_6995, Slonczewski2007_169}. The STT results from the transfer of angular momentum and provides the opportunity to excite high-frequency magnetization oscillations in nanomagnets by applied direct or alternating (microwave) current          ~\cite{Kiselev2003_380}-\cite{Fang2016_11259}. Furthermore, spin-polarized charge currents with sufficiently high densities lead to magnetization switching in metallic pillars ~\cite{Myers1999_867, Katine2000_3149} and magnetic tunnel junctions (MTJs) ~\cite{Huai2004_3118}-\cite{Sato2011_042501}. Such current-induced switching serves as a mechanism for data writing in magnetic random access memories utilizing the STT effect (STT-MRAMs) ~\cite{Ikeda2007_991}-\cite{Ando2014_172607}, while the magnetization precession driven by direct currents in spin-torque nanoscale oscillators (STNOs) creates microwave voltages, which makes STNOs potentially useful as frequency-tunable microwave sources and detectors ~\cite{Bonetti2009_102507}-\cite{Fang2016_11259}.

In ferromagnetic nanostructures comprising insulating interlayers, the electric field created in the insulator adjacent to the metallic ferromagnet may significantly affect the magnetic anisotropy of the latter. Such voltage-controlled magnetic anisotropy (VCMA) results from the penetration of electric field into an atomically thin surface layer of the ferromagnetic metal, which modifies the interfacial magnetic anisotropy ~\cite{Weisheit2007_349}-\cite{Alzate2014_112410}. The presence of VCMA renders possible to induce the magnetization precession in ferromagnetic nanostructures by microwave voltages ~\cite{Nozaki2012_491}-\cite{Miura2017_42511}. It is also shown that the application of dc voltage to the ferromagnetic nanolayer possessing VCMA can lead to a spin reorientation transition (SRT) ~\cite{Duan2008_122905}-\cite{Pertsev2014_024426}. Moreover, precessional 180$^{\circ}$ magnetization switching using electric field pulses has been demonstrated experimentally ~\cite{Shiota2012_39, Grezes2016_012403}. In addition, the voltage dependence of the interfacial magnetic anisotropy in CoFeB/MgO/CoFeB tunnel junctions may greatly reduce the critical current density required for the STT-driven magnetization reversal ~\cite{Wang2011_64, Pertsev2013_02757}. 

Importantly, magnetization precession in a ferromagnetic layer gives rise to spin pumping into adjoining normal metal or semiconductor ~\cite{Tserkovnyak2002_117601}-\cite{Ando2011_655}. In this paper, we theoretically study the magnetization dynamics driven by a direct current applied to the Co$_{20}$Fe$_{60}$B$_{20}$/MgO/Co$_{20}$Fe$_{60}$B$_{20}$ tunnel junction and calculate the time-dependent spin current generated in the Au overlayer. The magnetization evolution in the free Co$_{20}$Fe$_{60}$B$_{20}$ layer is determined by solving numerically the Landau-Lifshitz-Gilbert-Slonczewski (LLGS) equation, which accounts for the STT created by a spin-polarized tunnel current and for the VCMA associated with the Co$_{20}$Fe$_{60}$B$_{20}|$MgO interface. A range of current densities is revealed, within which a steady-state magnetization precession is generated in the free Co$_{20}$Fe$_{60}$B$_{20}$ layer. For this "{\it precession window}", frequencies and trajectories of magnetization oscillations are determined and used to calculate the time-dependent spin current created in the Au overlayer. Our calculations are distinguished by the account of both the spin polarization of the charge current and the precession-driven spin pumping as well as the contribution of the latter to the damping of magnetization dynamics. Finally, we solve coupled drift-diffusion equations for charge and spin currents to determine the spatial distribution of the electric potential in the Co$_{20}$Fe$_{60}$B$_{20}$/Au bilayer.

\section{\label{sec:two}CURRENT-DRIVEN MAGNETIZATION DYNAMICS}
\begin{figure}[h!]
\centering
\includegraphics[width=0.73\linewidth]{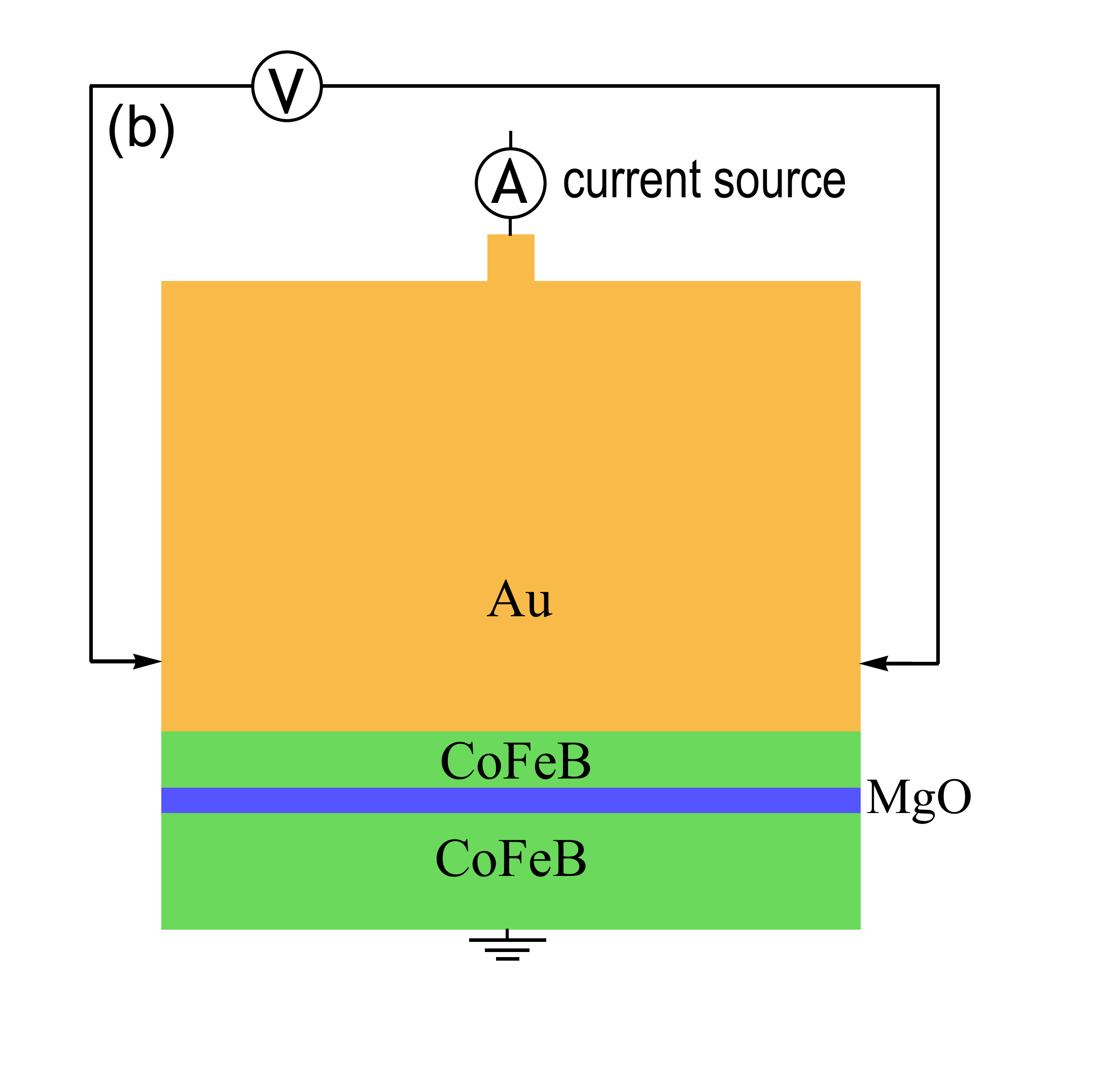}
\caption[]{
\justifying
Schematic representation of CoFeB/MgO/CoFeB/Au tunneling heterostructure connected to a constant-current source. (a) CoFeB/MgO/CoFeB tunnel junction comprising ultrathin free layer having almost perpendicular-to-plane magnetization \textbf{M} and thick pinned layer with in-plane magnetization \textbf{M}$_p$. (b) Measurement of the transverse voltage signal generated by the Au overlayer.}
\label{image1}
\end{figure}

We consider an MTJ comprising an ultrathin free layer with the thickness $t_f$ smaller than the critical thickness $t_{\textrm{\tiny{SRT}}}$, below which it acquires a perpendicular magnetic anisotropy ~\cite{Kanai2012_122403, Ikeda2010_721}. The thickness $t_p$ of the pinned layer is taken to be larger than $t_{\textrm{\tiny{SRT}}}$ so that the pinned magnetization {\bf M}$_p$ has an in-plane orientation (Fig.~\ref{image1}). Both layers are assumed to be homogeneously magnetized, and the current flowing through the tunnel barrier is regarded uniform. 

\begin{figure}[h!]
\centering
\includegraphics[width=0.95\linewidth]{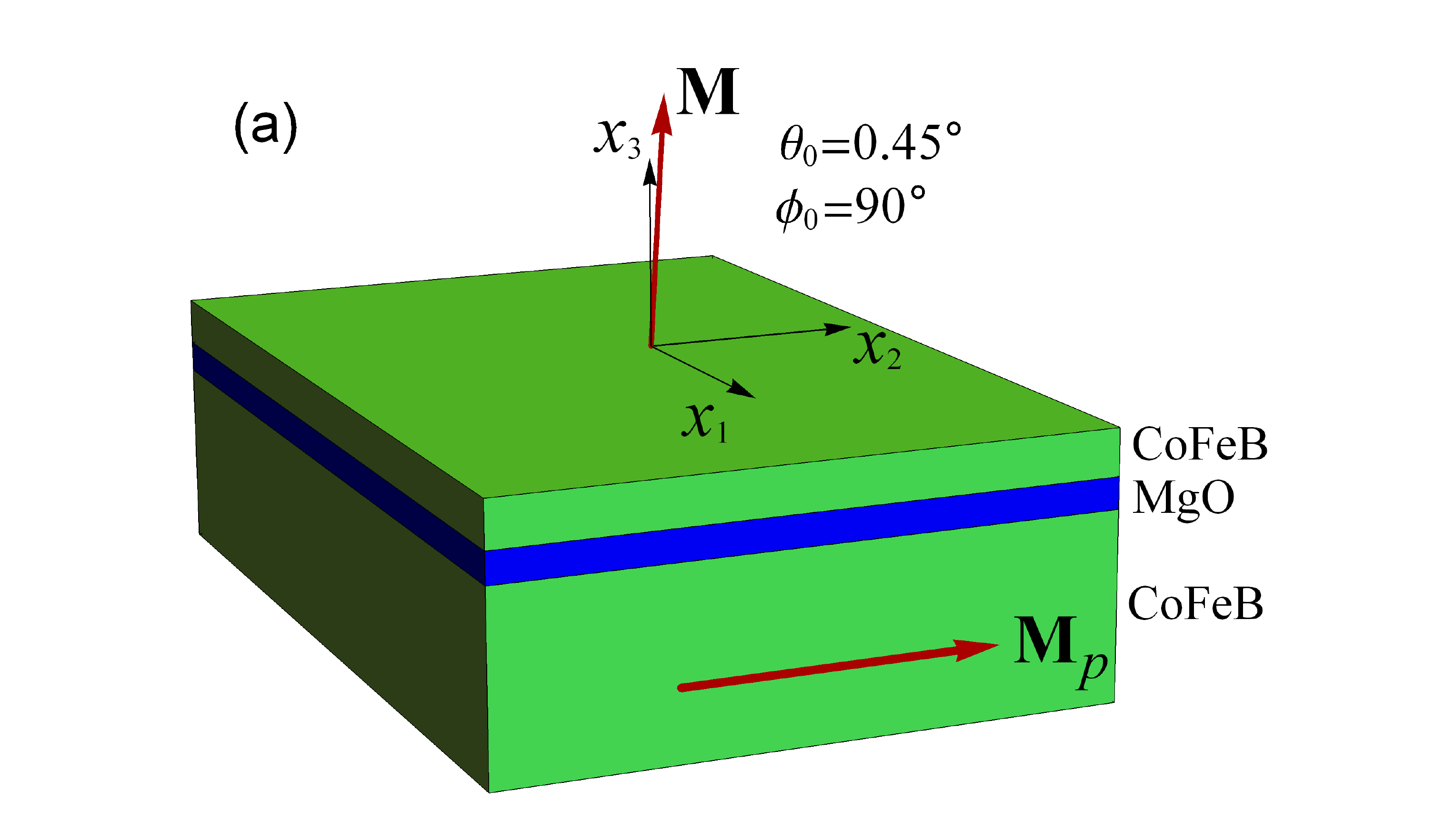}
\end{figure}

\noindent
To describe the dynamics of the free-layer magnetization {\bf M}$(t)$, we employ the macrospin approximation, which is well-suited for magnetic layers with nanoscale in-plane dimensions. Since the magnetization magnitude $|\bf {M}|$ = $M_s$ at a fixed temperature much lower than the Curie temperature can be considered a constant quantity, the LLGS equation may be reformulated for the unit vector ${\bf m} = {\bf M}/M_s$ ~\cite{Brataas2012_372} and written as

\begin{equation}
\label{eq1}
\begin{gathered}
    \frac{d{\bf m}}{dt} = -\gamma \mu_0 {\bf m} \times {\bf H}_{\textrm{\tiny{eff}}} + \alpha {\bf m} \times \frac{d{\bf m}}{dt} \\ +
    \frac{\tau_{\textrm{\tiny{STT}}}}{M_s} {\bf m}\times({\bf m} \times {\bf m}_p),
\end{gathered}
\end{equation}

\noindent
where $\gamma > 0$ is the electron's gyromagnetic ratio, $\mu_0$ is the permeability of vacuum, $\alpha$ is the Gilbert damping parameter, and ${\bf{H}_{\textrm{\tiny{eff}}}}$ is the effective field acting on the magnetization. In Eq. \hyperref[eq1]{(1)}, the last term takes into account the STT proportional to the current density $J$ in the free layer, whereas the field-like torque is disregarded because it does not affect the magnetic dynamics qualitatively ~\cite{Fang2016_11259, Zhu2012_197203}. For symmetric MTJs, the theory gives $\tau_{\textrm{\tiny{STT}}} = (\gamma \hbar/ 2 e)(J / t_f) \eta / (1+\eta^2 {\bf m} \cdot {\bf m}_p)$, where $e$ is the elementary (positive) charge, $\hbar$ is the reduced Planck constant, $\eta = \sqrt{(G_\textrm{\tiny{P}}-G_\textrm{\tiny{AP}})/(G_\textrm{\tiny{P}}+G_\textrm{\tiny{AP}})}$ and $G_\textrm{\tiny{P}}$ and $G_\textrm{\tiny{AP}}$ are the MTJ conductances per unit area in the states with parallel and antiparallel electrode magnetizations, respectively ~\cite{Slonczewski2007_169}. Since we consider the MTJ connected to a constant-current source, the voltage drop $V = J/G$ across the tunnel barrier depends on the junction's conductance $G = G_\textrm{\tiny{P}}(1+\eta^2 {\bf m} \cdot {\bf m}_p)/(1+\eta^2)$, which leads to a non-sinusoidal dependence of the STT on the angle between {\bf m} and ${\bf m}_p$. The effective field involved in Eq. \hyperref[eq1]{(1)} is defined by the relation ${\bf H}_\textrm{\tiny{eff}}=-(\mu_0 M_s)^{-1}\partial F / \partial {\bf m}$, where $F({\bf m})$ is the Helmholtz free energy density of ferromagnetic layer. For a homogeneously magnetized unstrained free layer made of cubic ferromagnet, the magnetization-dependent part $\Delta F({\bf m})$ of the effective volumetric energy density may be approximated by the polynomial

\begin{equation}
\label{eq2}
\begin{gathered}
    \Delta F = K_1(m_1^2m_2^2+m_1^2m_3^2+m_2^2m_3^2) + K_2 m_1^2m_2^2m_3^2 
    \\+\frac{K_s}{t_f}m_3^2 
    - \frac{U_\textrm{\tiny{IEC}}}{t_f} (m_1 m_1^p+m_2 m_2^p+m_3 m_3^p)\\
    +\frac{1}{2}\mu_0 M_s^2(N_{11}m_1^2+2N_{12}m_1m_2\\+N_{22}m_2^2+N_{33}m_3^2)\\
    -\mu_0 M_s (H_1 m_1+H_2 m_2+H_3 m_3)
\end{gathered}
\end{equation}

\noindent 
where $m_i$ ($i$ = 1, 2, 3) are the direction cosines of {\bf M} in the crystallographic reference frame with the $x_3$ axis orthogonal to the layer surfaces, $K_1$ and $K_2$ are the coefficients of the fourth- and sixth-order terms defining the cubic magnetocrystalline anisotropy, $K_s$ is the parameter characterizing the total specific energy of two interfaces (Co$_{20}$Fe$_{60}$B$_{20}|$MgO and Co$_{20}$Fe$_{60}$B$_{20}|$Au in our case), $U_\textrm{\tiny{IEC}}$ is the energy of interlayer exchange coupling (IEC) with the pinned layer (per unit area), $N_{ij}$ are the demagnetizing factors ($N_{13}$ and $N_{23}$ are negligible at in-plane dimensions $L_1$, $L_2 >> t_f$), and {\bf H} is the average magnetic field acting on the free layer. Since the magnetic anisotropy associated with the Co$_{20}$Fe$_{60}$B$_{20}|$MgO interface depends on the electric field $E_3$ in MgO ~\cite{Kanai2012_122403, Alzate2014_112410}, the coefficient $K_s$ is a function of the current density $J$. Using a linear approximation for the dependence $K_s(E_3)$ supported by first-principles calculations ~\cite{Niranjan2010_222504} and experimental data ~\cite{Alzate2014_112410}, we arrive at the relation $K_s = K_s^0+k_s V/t_b = K_s^0+k_s J/(G t_b)$, where $K_s^0=K_s(E_3=0)$, $k_s=\partial K_s / \partial E_3$ is the electric-field sensitivity of $K_s$, and $V$ is the voltage drop across the MgO layer of thickness $t_b$, which is caused by the tunnel current flowing through the junction with the conductance $G$ per unit area. 

The numerical integration of Eq. \hyperref[eq1]{(1)} was realized with the aid of the projective Euler scheme, where the condition $|{\bf m}| = 1$ is satisfied automatically. A fixed integration step $\delta t = 0.5$ fs was used in our computations. The effective field ${\bf H}_\textrm{\tiny{eff}}$ was determined from Eq. \hyperref[eq2]{(2)} under the assumption of negligible total magnetic field {\bf H} acting on the free layer, which is justified by the absence of external magnetic sources and zero mean value of the current-induced Oersted field. Since in the considered heterostructure the magnetization dynamics in the free layer leads to the spin pumping into adjacent nonmagnetic layer, the parameters $\gamma$ and $\alpha$ involved in Eq. \hyperref[eq1]{(1)} were renormalized as ~\cite{Tserkovnyak2002_117601}

\begin{equation}
\label{eq3}
\begin{gathered}
\alpha = \frac{\gamma}{\gamma_0} \bigg\{  \alpha_0 + \frac{g_L \mu_B}{4\pi M_s t_f} \textrm{Re}[g_{\uparrow \downarrow}^r]  \bigg\},\\
\frac{1}{\gamma} = \frac{1}{\gamma_0} \bigg\{ 1+\frac{g_L \mu_B}{4 \pi M_s t_f} \textrm{Im}[g_{\uparrow \downarrow}^r] \bigg\},
\end{gathered}
\end{equation}

\noindent
where $\gamma_0$ and $\alpha_0$ denote the bulk values of $\gamma$ and $\alpha$, $g_L$ is the Land\'e factor, $\mu_B$ is the Bohr magneton, and $g_{\uparrow \downarrow}^r$ is the complex reflection spin-mixing conductance per unit area of the ferromagnet/normal metal contact ~\cite{Zwierzycki2005_064420}. The Gilbert parameter $\alpha_0$ was regarded as a constant quantity, because numerical estimates show that the dependence of $\alpha_0$ on the power of magnetization precession ~\cite{Slavin2009_1875} is negligible in our case.

\begin{figure*}[t]
\centering
\begin{minipage}[h]{0.45\linewidth}
\includegraphics[width=0.9\linewidth]{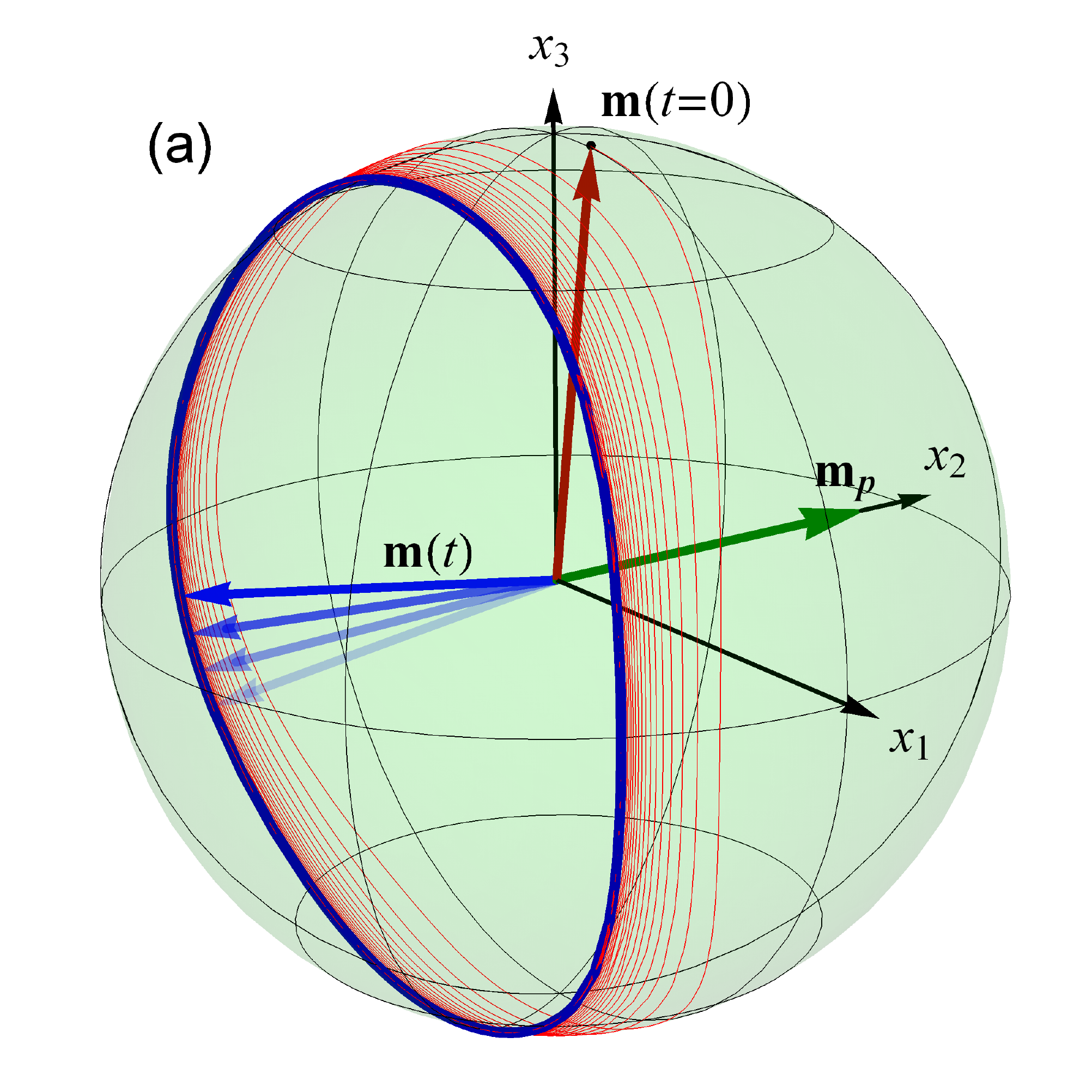}
\end{minipage}
\hfill
\begin{minipage}[h]{0.49\linewidth}
\includegraphics[width=0.9\linewidth]{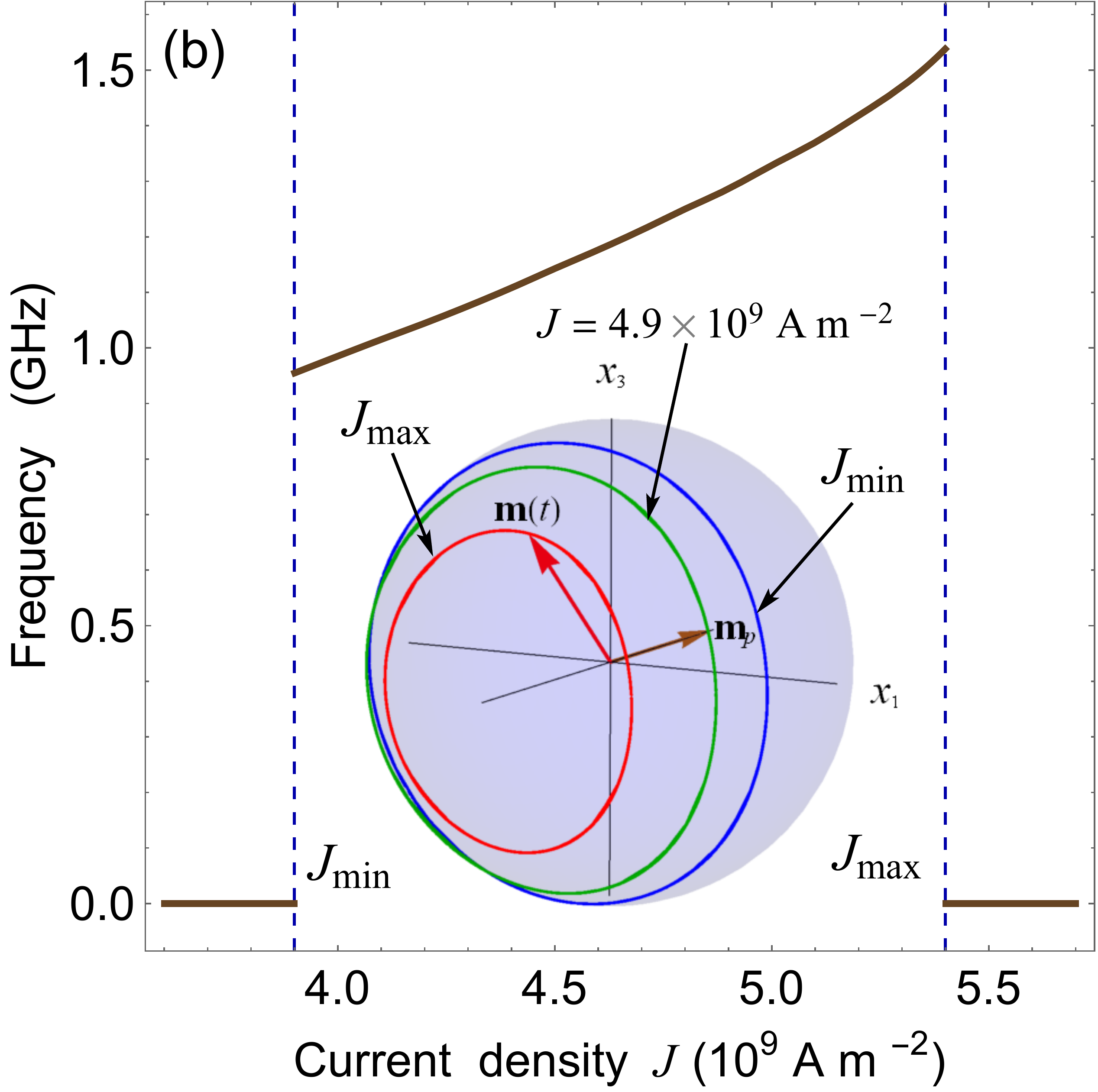}
\end{minipage}
\caption[]{
\justifying
Current-driven magnetization dynamics in the free Co$_{20}$Fe$_{60}$B$_{20}$ layer. (a) Trajectory of the end of the unit vector \textbf{m} after the destabilization caused by the critical current density $J_\textrm{\tiny{min}}$. (b) Frequency \textit{f} of the steady-state magnetization precession as a function of the tunnel-current density \textit{J}. The inset shows the magnetization trajectories obtained at $J = J_\textrm{\tiny{min}}$ (blue curve), $J = 4.9 \times 10^9$ A m$^{-2}$ (green curve), and $J = J_\textrm{\tiny{max}}$ (red curve).}
\label{image2}
\end{figure*}

The numerical calculations were performed for the Co$_{20}$Fe$_{60}$B$_{20}$/MgO/Co$_{20}$Fe$_{60}$B$_{20}$ junction with the barrier and electrode thicknesses equal to $t_b = 1.1$ nm, $t_f = 1.73$ nm, and $t_p = 5$ nm. A rectangular in-plane shape and nanoscale dimensions $L_1 = 400$ nm and $L_2 = 40$ nm were chosen for the free layer. The demagnetizing factors of such ferromagnetic layer, calculated from the available analytic formulae  ~\cite{Aharoni1998_3432}, were found to be $N_{11} = 0.0059$, $N_{22} = 0.0626$, $N_{12} = 0$, and $N_{33} = 0.9315$. A high in-plane aspect ratio $L_1 / L_2 = 10$ was given to the free layer in order to make energetically more favorable the magnetization orientations in the plane perpendicular to the pinned magnetization ${\bf M}_p$, which enhances the STT acting on {\bf M}. The pinned layer was assumed to have a large area ensuring negligible contribution of the magnetostatic interlayer interaction to the free-layer energy $\Delta F$ in comparison with that of the IEC defined by the relation $U_\textrm{\tiny{IEC}} \approx 5.78 \textrm{exp}(-7.43\times 10^9 t_b ~\mathrm{m}^{-1})$ mJ m$^{-2}$ ~\cite{Skowronski2010_093917}. The saturation magnetization $M_s = 1.13 \times 10^6$ A m$^{-1}$  ~\cite{Lee2011_123910} and the Gilbert damping constant $\alpha_0 = 0.01$ ~\cite{Ikeda2010_721} were assigned to the Co$_{20}$Fe$_{60}$B$_{20}$ free layer, while its magnetocrystalline anisotropy was described using the coefficients $K_1 = 5 \times 10^3$ J m$^{-3}$ ~\cite{Hall1960_S157} and $K_2 = 50$ J m$^{-3}$  ~\cite{Zhu2012_197203}. To quantify the VCMA associated with the Co$_{20}$Fe$_{60}$B$_{20}|$MgO interface, we used the measured parameters $K_s^0 = -1.3 \times 10^{-3}$ J m$^{-2}$  ~\cite{Ikeda2010_721} and $k_s = 37$ fJ V$^{-1}$ m$^{-1}$  ~\cite{Zhu2012_197203}. The junction's conductance $G_\textrm{\tiny{P}}$ at the chosen MgO thickness was taken to be $8.125 \times 10^9$ S m$^{-2}$ ~\cite{Tsunekawa2005_FB-05}, and we used typical asymmetry parameter $\eta = 0.57$ ~\cite{Sato2011_042501, Alzate2014_112410} which yields the tunneling magnetoresistance ratio TMR $ = (G_\textrm{\tiny{P}} - G_\textrm{\tiny{AP}}) / G_\textrm{\tiny{AP}} \cong 96 \%$.

The numerical calculations started with the determination of the equilibrium magnetization orientation in the free Co$_{20}$Fe$_{60}$B$_{20}$ layer at zero applied current. It was found that the initial energy landscape $\Delta F (\phi_0, \theta_0)$ has only two minima, which correspond to almost perpendicular-to-plane directions of the free-layer magnetization \textbf{M}. Owing to the IEC with the in-plane magnetized pinned Co$_{20}$Fe$_{60}$B$_{20}$ layer, the magnetization \textbf{M} slightly deviates from the perpendicular-to-plane orientation, tilting towards the pinned magnetization \textbf{M}$_p$ oriented along the $x_2$ axis ($\phi_0 = 90^{\circ}$, $\theta_0 = 0.45^{\circ}$ or $179.55^{\circ}$, see Fig.~\ref{image1}). The energy barrier for the coherent magnetization switching at room temperature $T_r$ is about $60 k_B T_r$, where $k_B$ is the Boltzmann constant. Importantly, the perpendicular magnetic anisotropy is sufficient to prevent the coexistence of metastable states with an in-plane orientation of \textbf{M}, which otherwise could temporarily show up due to thermal fluctuations. 

The application of a small current to the MTJ modifies the equilibrium magnetization orientation because the interfacial magnetic anisotropy changes due to a voltage drop $V = J/G$ across the barrier and a nonzero $\tau_\textrm{\tiny{STT}}(J)$ appears in Eq. \hyperref[eq1]{(1)}. The simulations showed that at $J < 0$ the magnetization \textbf{M} progressively rotates towards the PP direction with increasing current, remaining stable up to very high densities $|J| < 10^{10}$ A m$^{-2}$. On the contrary, the deviation of \textbf{M} from the PP direction increases when a positive current is applied to the MTJ ($J > 0$), reaching $\theta = 7.54^{\circ}$ just below the critical density $J_\textrm{\tiny{min}} \cong 3.9 \times 10^9$ A m$^{-2}$ at which the magnetization dynamics arises. Remarkably, the predicted value of $J_\textrm{\tiny{min}}$ falls into the range of lowest critical current densities $|J_\textrm{\tiny{min}}(t_f)| = (1.2-5.4) \times 10^9$ A m$^{-2}$ measured experimentally up to date ~\cite{Zeng2013_1426}. Therefore, we focus below on the magnetic dynamics induced by positive applied currents, which correspond to the tunneling of electrons from the free layer into the pinned one.

Figure~\ref{image2} (a) shows the trajectory of the end of the unit vector ${\bf m} = {\bf M} / M_s$ after the destabilization caused by the positive current with the critical density $J_\textrm{\tiny{min}}$. Remarkably, the free Co$_{20}$Fe$_{60}$B$_{20}$ layer experiences a {\it dynamic SRT}, at which the static magnetic state with almost PP orientation of \textbf{m} transforms into large-angle magnetization precession around in-plane (IP) direction antiparallel to the pinned magnetization \textbf{M}$_p$. The appearance of such electrically driven SRT can be attributed to the proximity of the free-layer thickness $t_f = 1.73$ nm to the critical thickness $t_\textrm{\tiny{SRT}} = 1.745$ nm, at which the size-induced SRT should take place in the considered MTJ at $J = 0$. Indeed, the change $\Delta K_s = k_s J_\textrm{\tiny{min}} / (G t_b)$ in VCMA promotes voltage-driven SRT to the IP magnetization orientation parallel to the $x_1$ axis, while the STT gives rise to the precession of \textbf{m}. The proximity to the thickness-induced SRT also explains very large precession amplitude at $J_\textrm{\tiny{min}}$. With increasing current density $J > J_\textrm{\tiny{min}}$, the frequency of steady-state magnetization precession rises, whereas its amplitude becomes smaller [Fig.~\ref{image2}(b)]. The precession frequency $\nu$ ranges from 0.95 GHz at $J_\textrm{\tiny{min}}$ to 1.54 GHz at the maximal density $J_\textrm{\tiny{max}} = 5.4 \times 10^9$ A m$^{-2}$ above which the precession disappears ~\cite{Note1}. Owing to strong STT, the free-layer magnetization stabilizes at $J > J_\textrm{\tiny{max}}$ along the direction antiparallel to the magnetization of the pinned Co$_{20}$Fe$_{60}$B$_{20}$ layer.

\section{\label{sec:three}SPIN AND CHARGE CURRENTS IN NORMAL-METAL OVERLAYER}

\begin{figure*}[t]
\centering
\begin{minipage}[h]{0.49\linewidth}
\includegraphics[width=0.9\linewidth]{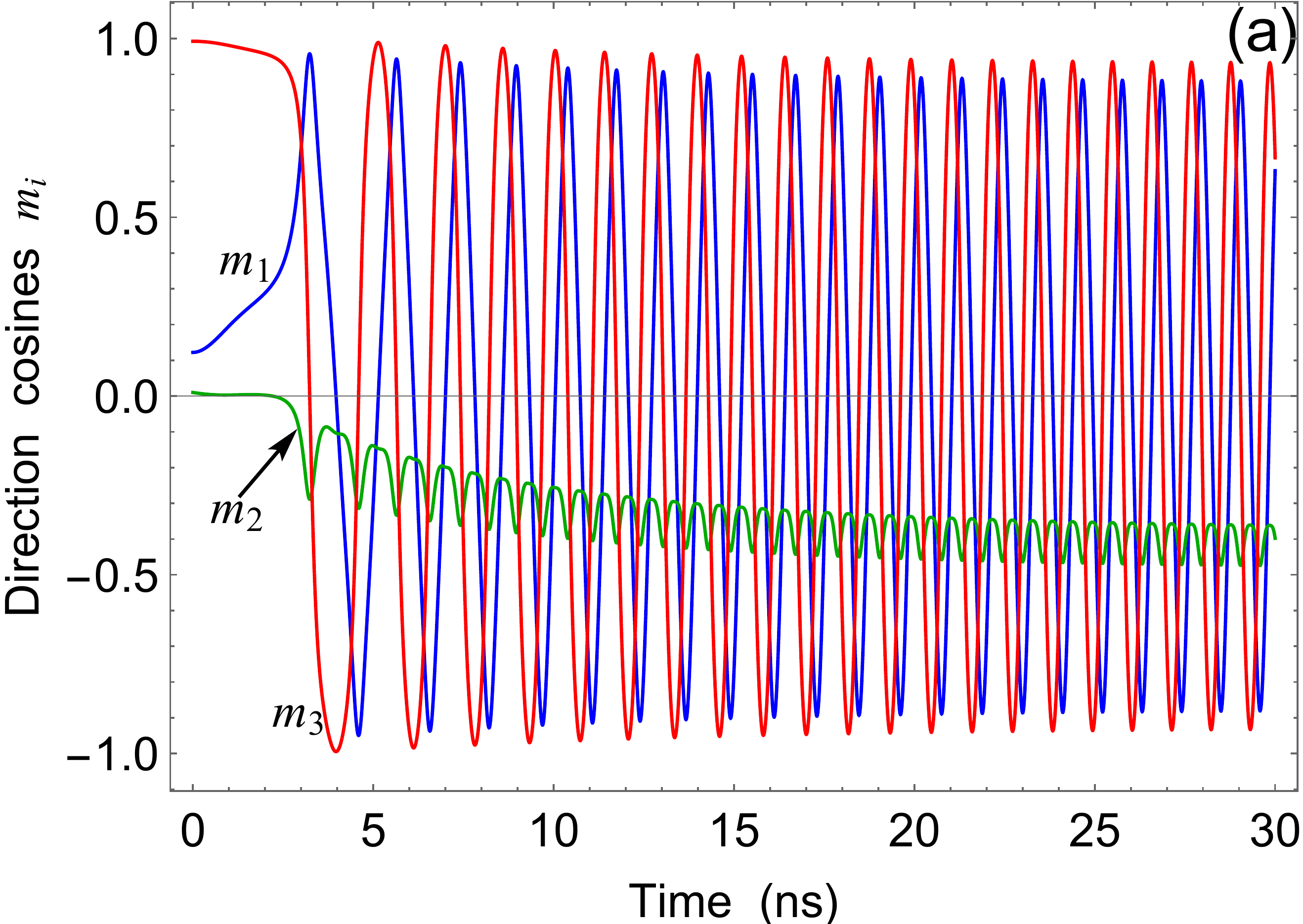}
\end{minipage}
\hfill
\begin{minipage}[h]{0.495\linewidth}
\includegraphics[width=0.9\linewidth]{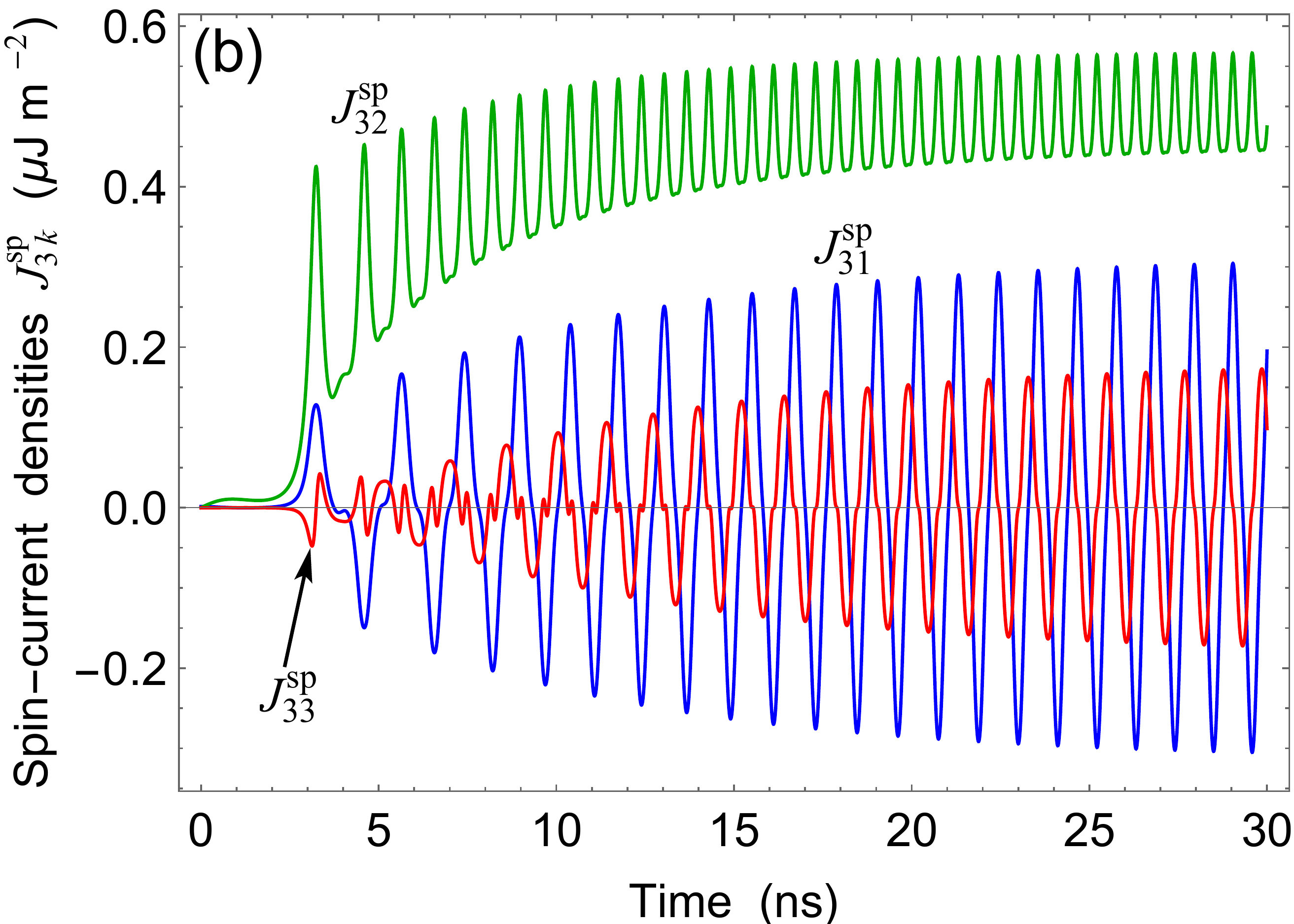}
\end{minipage}
\caption[]{
\justifying
Spin pumping into the Au overlayer generated by the magnetization dynamics appearing in the free Co$_{20}$Fe$_{60}$B$_{20}$ layer at the critical charge-current density $J_\textrm{\tiny{min}}$. (a) Time dependences of the direction cosines $m_i$ of the free-layer magnetization. (b) Temporal evolutions of the spin-current densities $J_{3k}^\textrm{\tiny{sp}}$ at the Au$|$Co$_{20}$Fe$_{60}$B$_{20}$ interface.}
\label{image3}
\end{figure*}

The electrically induced magnetic dynamics in the free Co$_{20}$Fe$_{60}$B$_{20}$ layer should lead to the spin pumping into the Au overlayer. The spin-current density can be specified by a tensor ${\bf J}_s$ characterizing both the direction of spin flow defined by the unit vector ${\bf e}_s$ and the orientation of spin polarization ~\cite{Dyakonov1971_459}. Since the Co$_{20}$Fe$_{60}$B$_{20}$ thickness is well above a few monolayer range, the imaginary part of the reflection spin-mixing conductance $g_{\uparrow \downarrow}^r$ and the transmission spin mixing conductance $g_{\uparrow \downarrow}^t$ are negligible. Therefore, the pumped spin-current density ${\bf J}_\textrm{\tiny{sp}}$ in the vicinity of the Co$_{20}$Fe$_{60}$B$_{20}|$Au interface can be calculated from the approximate relation ${\bf e}_s \cdot {\bf J}_\textrm{\tiny{sp}} \cong (\hbar / 4\pi)\textrm{Re}[g_{\uparrow \downarrow}^r]{\bf m} \times d{\bf m} / dt$ ~\cite{Zwierzycki2005_064420}. Adopting for the Co$_{20}$Fe$_{60}$B$_{20}|$Au interface the theoretical estimate $(e^2 / h)\textrm{Re}[g_{\uparrow \downarrow}^r] \approx 4.66 \times 10^{14}$ $\Omega^{-1}$ m$^{-2}$ obtained for the Fe$|$Au one ~\cite{Zwierzycki2005_064420}, we calculated the spin current pumped into Au during the magnetization precession in the free layer. Figure~\ref{image3} shows representative time dependences of the nonzero spin-current densities $J_{3k}^\textrm{\tiny{sp}}(t)$ ($k = $ 1, 2, 3), which correspond to the magnetization dynamics $\textbf{m}(t)$ appearing at the critical charge-current density $J_\textrm{\tiny{min}}$. Interestingly, $J_{32}^\textrm{\tiny{sp}}$ contains significant dc and ac components, whereas $J_{31}^\textrm{\tiny{sp}}$ and $J_{33}^\textrm{\tiny{sp}}$ are dominated by the ac component. In the steady-state regime, $J_{32}^\textrm{\tiny{sp}}$ oscillates with the frequency $2\nu$, which is two times higher than the precession frequency $\nu$ due to similar oscillations of the direction cosine $m_2$. 

\begin{figure*}[t]
\centering
\begin{minipage}[h]{0.454\linewidth}
\includegraphics[width=0.9\linewidth]{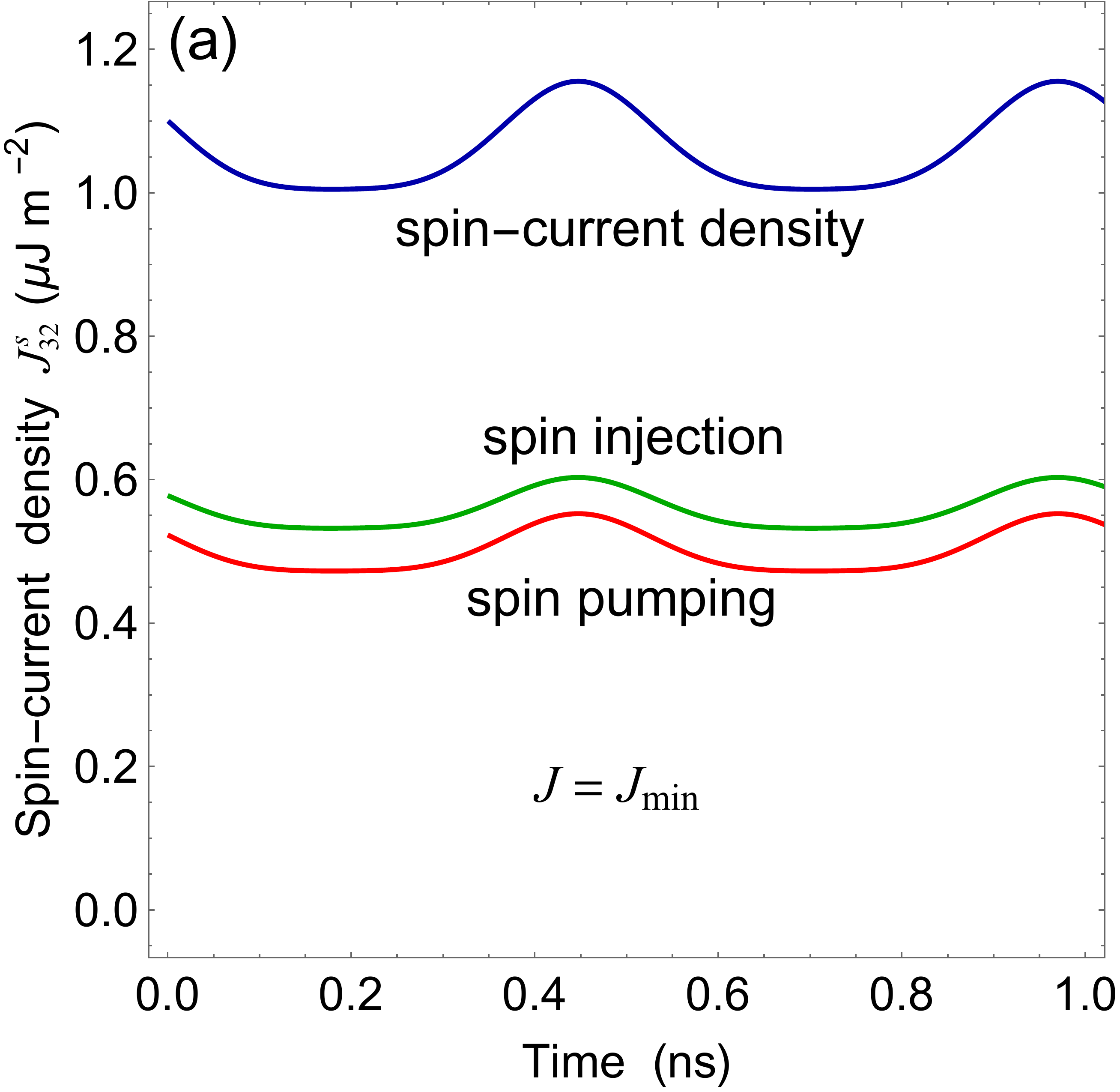}
\end{minipage}
\hfill
\begin{minipage}[h]{0.45\linewidth}
\includegraphics[width=0.9\linewidth]{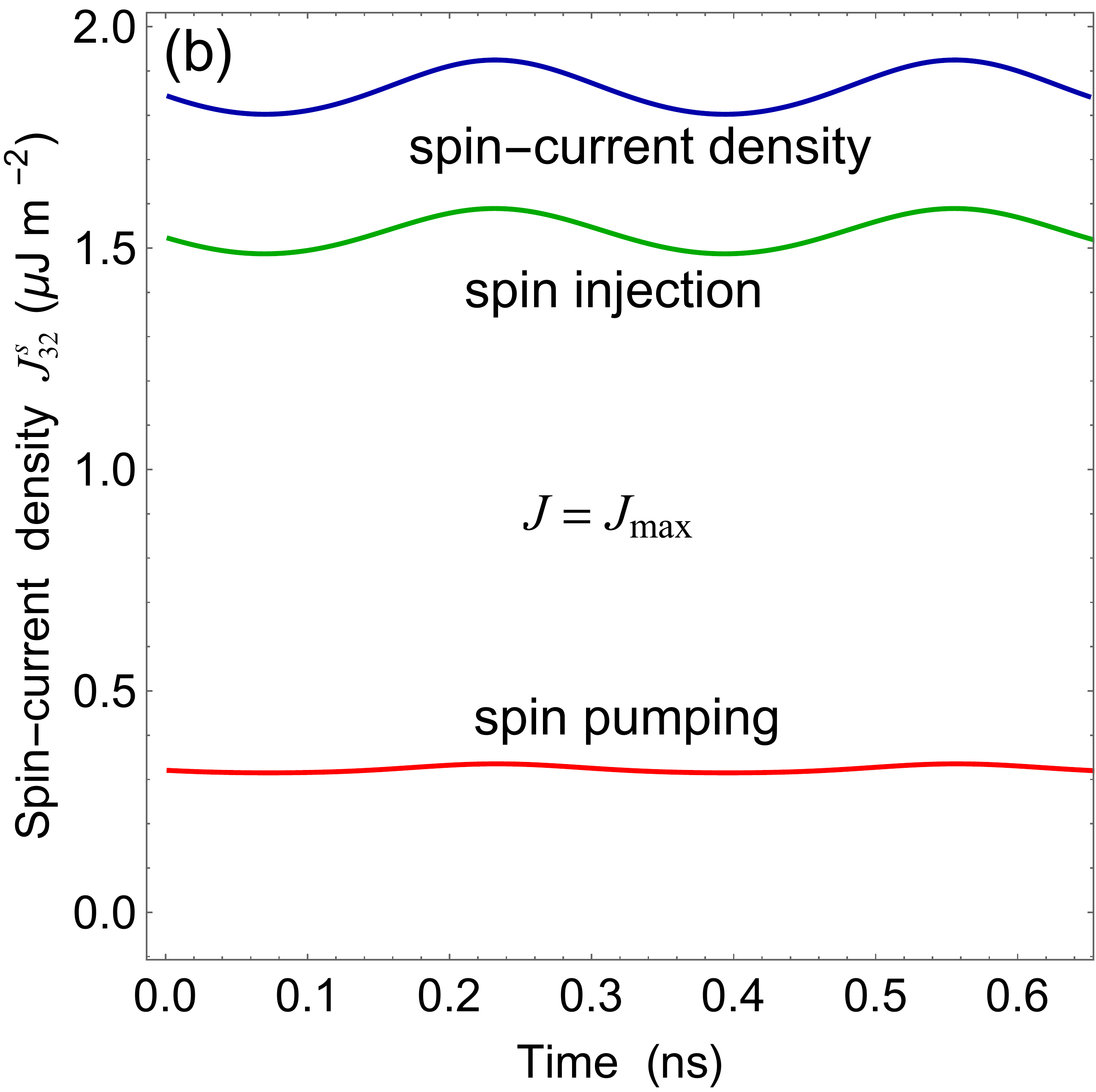}
\end{minipage}
\caption[]{
\justifying
Total spin-current density $J_{32}^s$ generated near the Au$|$Co$_{20}$Fe$_{60}$B$_{20}$ interface at the critical charge-current densities $J_\textrm{\tiny{min}}$ (a) and $J_\textrm{\tiny{max}}$ (b) defining the precession window in the free Co$_{20}$Fe$_{60}$B$_{20}$ layer. Contributions of spin-polarized charge current and precession-induced spin pumping are shown by green and red lines, respectively.}
\label{image4}
\end{figure*}

Taking into account the spin polarization of the charge current governed by the free-layer magnetization $\textbf{M}(t)$, we calculated the total spin-current density $\textbf{J}_s = \textbf{J}_\textrm{\tiny{sp}} + \textbf{J}_\textrm{\tiny{sc}}$ at the Co$_{20}$Fe$_{60}$B$_{20}|$Au interface. Figure~\ref{image4} shows time dependences of the most interesting component $J_{32}^s(t)$ evaluated at critical charge-current densities $J_\textrm{\tiny{min}}$ and $J_\textrm{\tiny{max}}$. Remarkably, the contributions of spin-polarized charge current ($J_{32}^\textrm{\tiny{sc}}$) and precession-induced spin pumping ($J_{32}^\textrm{\tiny{sp}}$) have the same sign and phase. At $J = J_\textrm{\tiny{min}}$, both $J_{32}^\textrm{\tiny{sc}}$ and $J_{32}^\textrm{\tiny{sp}}$ exhibit non-sinusoidal time dependences, whereas at $J = J_\textrm{\tiny{max}}$ the contribution $J_{32}^\textrm{\tiny{sc}}$ assumes almost sinusoidal shape while $J_{32}^\textrm{\tiny{sp}}$ becomes practically constant. 

\begin{figure*}[t]
\centering
\begin{minipage}[h]{0.488\linewidth}
\includegraphics[width=0.9\linewidth]{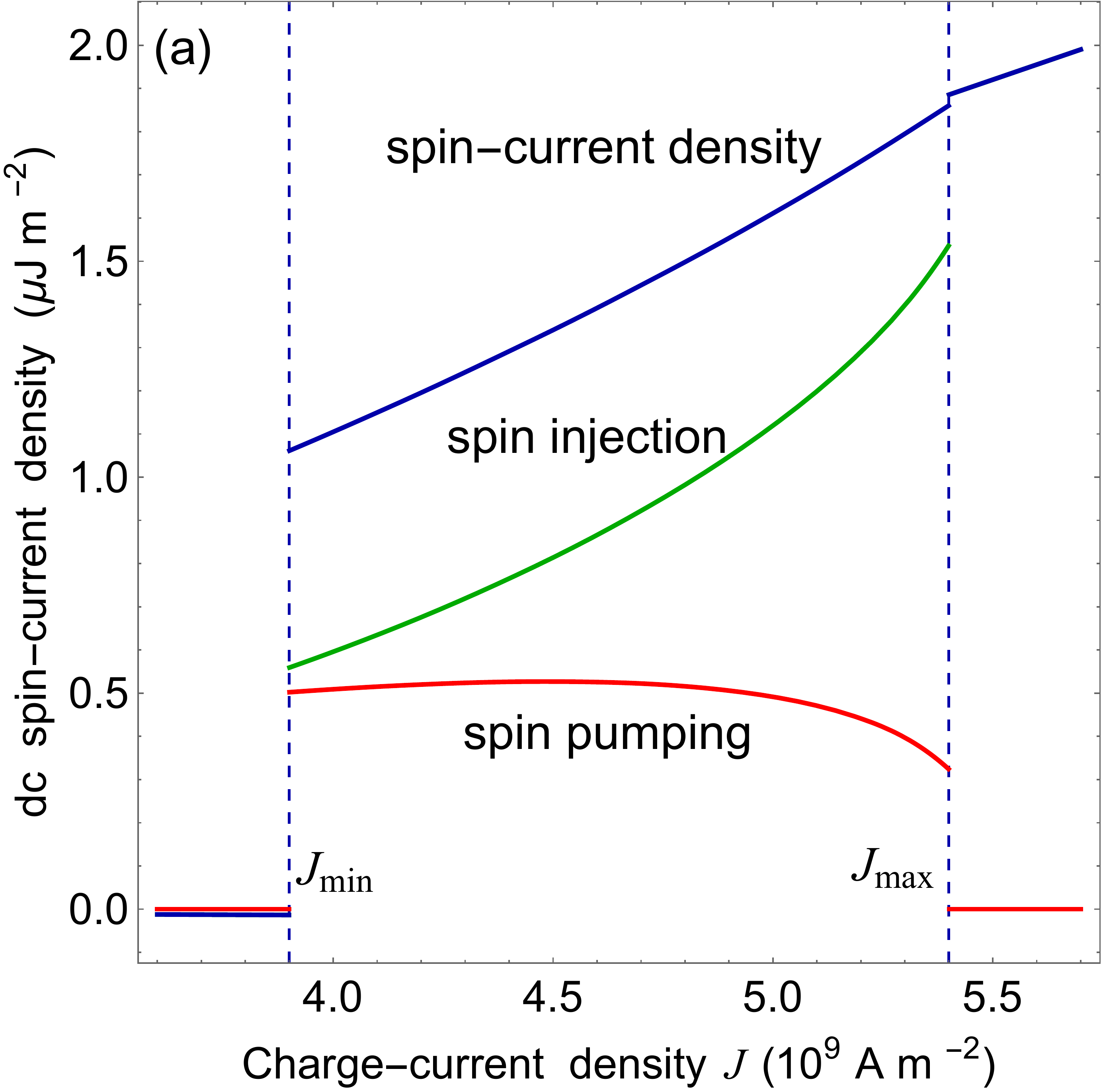}
\end{minipage}
\hfill
\begin{minipage}[h]{0.5\linewidth}
\includegraphics[width=0.9\linewidth]{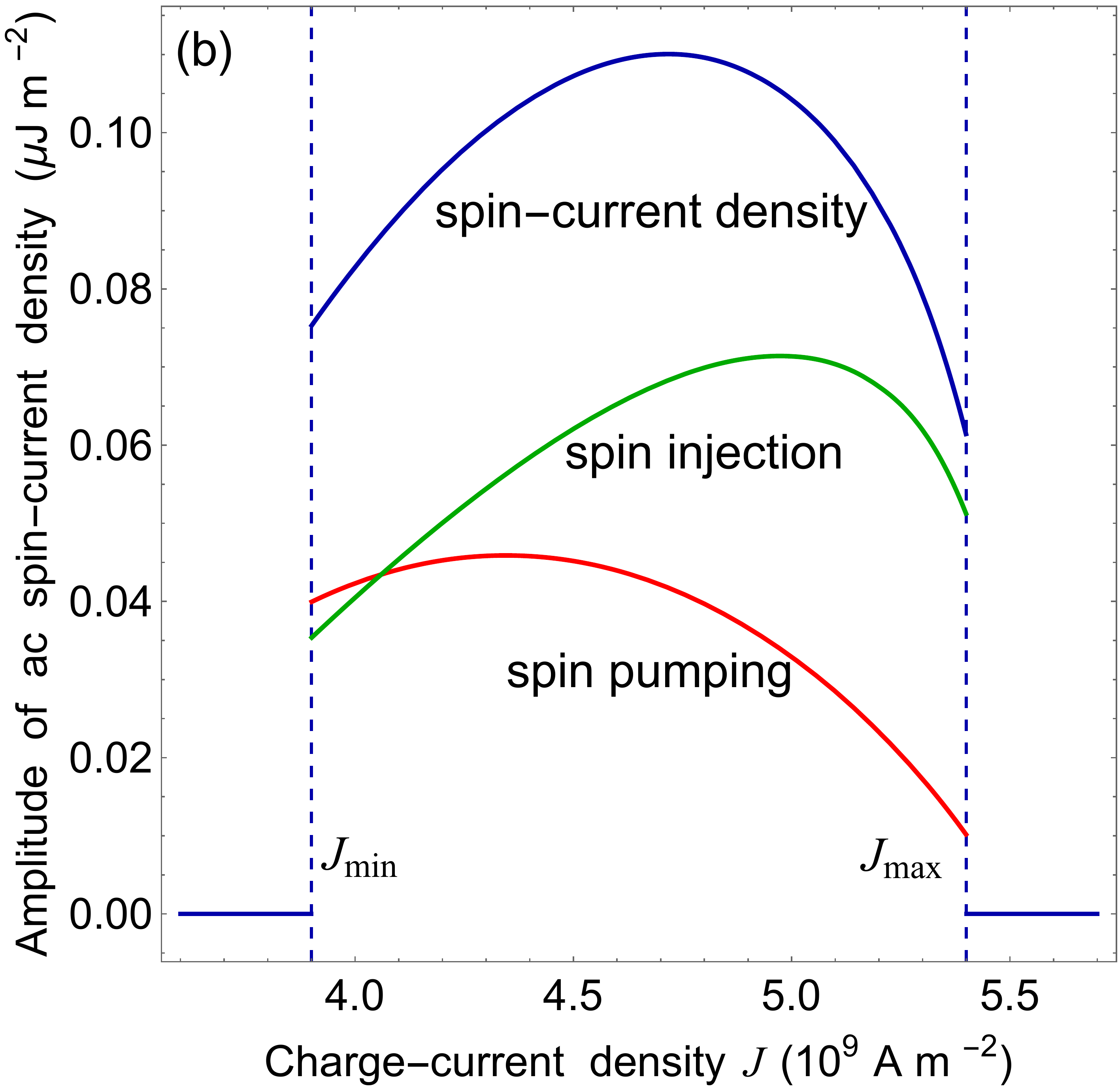}
\end{minipage}
\caption[]{ 
\justifying
Dependences of the total spin-current density $J_{32}^s$ generated near the Au$|$Co$_{20}$Fe$_{60}$B$_{20}$ interface on the density \textit{J} of the charge current flowing in the free Co$_{20}$Fe$_{60}$B$_{20}$ layer. Panel (a) shows the dc component $\left\langle J_{32}^s \right\rangle$ , while panel (b) presents the amplitude of the ac component of $J_{32}^s(t)$. Contributions of spin-polarized charge current and precession-induced spin pumping are shown by green and red lines, respectively.}
\label{image5}
\end{figure*}

The dc and ac components of the spin-current density $J_{32}^s(t)$ in the steady-state regime are plotted in Fig. ~\ref{image5} as a function of the charge-current density $J$. At $J < J_\textrm{\tiny{min}}$, the dc component $\left\langle J_{32}^s \right\rangle$ is small and negative due to the charge-current contribution $J_{32}^\textrm{\tiny{sc}}$, and the ac component is zero. Within the precession window $J_\textrm{\tiny{min}} < J < J_\textrm{\tiny{max}}$, $\left\langle J_{32}^s \right\rangle$ grows monotonically with increasing charge current owing to the significant rise of $J_{32}^\textrm{\tiny{sc}}$. In contrast, the amplitude of ac component becomes maximal at $J \cong 4.72 \times 10^9$ A m $^{-2}$ near the middle of the precession window.

Thus, the Co$_{20}$Fe$_{60}$B$_{20}$/MgO/Co$_{20}$Fe$_{60}$B$_{20}$ tunnel junction excited by a direct charge current can be employed for the generation of spin currents in normal metals ~\cite{Note2}. The power dissipation $W_\textrm{\tiny{min}} \cong J_\textrm{\tiny{min}}^2L_1L_2   \left\langle  G^{-1}  \right\rangle$ of such electrically driven spin injector is estimated to be below 40 $\mu$W which is a very small value for devices with sub-micrometer size ~\cite{Zeng2013_1426}. To evaluate the efficiency of the proposed spin injector, we calculated the electrical potential difference $\Delta V(x_3, t) = \phi(x_1 = L_1 / 2, x_3, t) - \phi(x_1 = -L_1 / 2, x_3, t)$ between the lateral sides of the Co$_{20}$Fe$_{60}$B$_{20}$Au bilayer. Owing to the inverse spin Hall effect, the spin flow in the normal metal creates such {\it transverse voltage signal}, which can be used to detect this flow experimentally ~\cite{Saitoh2006_182509}. 

To determine the distribution of the electric potential $\phi(\textbf{r}, t)$ in the Co$_{20}$Fe$_{60}$B$_{20}$/Au bilayer, we solved the coupled drift-diffusion equations ~\cite{Dyakonov1971_459, Takahashi2003_052409, Stiles2004_054408} for charge and spin currents flowing in the Co$_{20}$Fe$_{60}$B$_{20}$ and Au films. The continuity equations for the charge-current density \textbf{J} and the spin-current density $\textbf{J}_s$ have the form of $\nabla \cdot \textbf{J} = -\partial \rho / \partial t$ and $\hbar^{-1} \nabla \cdot \textbf{J}_s =-\partial \textbf{P} / \partial t - \textbf{P}/\tau_\textrm{\tiny{sf}}$, where $\rho$ is the charge density, \textbf{P} is the spin polarization density, and $\tau_\textrm{\tiny{sf}}$ is the spin-flip relaxation time. Since spatial variations in the electron concentration \textit{n} can be neglected for metals ~\cite{Stiles2004_054408}, explicit expressions for the densities \textbf{J} and $\textbf{J}_s$ reduce to

\begin{equation}
\label{eq4}
\begin{gathered}
{\bf J} = e \mu n {\bf E} + 2 \alpha_\textrm{\tiny{SH}} e \mu {\bf E} \times {\bf P} + 2 \alpha_\textrm{\tiny{SH}}e D \nabla \times {\bf P},
\end{gathered}
\end{equation}

\begin{equation}
\label{eq5}
\begin{gathered}
\frac{J_{ik}^s}{\hbar}=-\mu E_i P_k - D\frac{\partial P_k}{\partial x_i}+ 2 \alpha_\textrm{\tiny{SH}} \mu n \varepsilon_{ikl} E_l,
\end{gathered}
\end{equation}

\noindent
where \textbf{E} is the electric field, $\mu$ is the electron mobility, \textit{D} is the diffusion coefficient, $\alpha_\textrm{\tiny{SH}}$ is the spin Hall angle, $\varepsilon_{ikl}$ denote the components of the Levi-Civita tensor (\textit{i}, \textit{k}, \textit{l} = 1,2,3), and the Einstein summation convention is implied. In Eq. \hyperref[eq4]{(4)}, the second term describes the anomalous Hall effect characteristic of ferromagnetic metals, while the third term represents the inverse spin Hall effect. The first term in Eq. \hyperref[eq5]{(5)} gives the contribution of the spin-polarized charge current; the last term accounts for the spin Hall effect, which manifests itself in the current-induced spin accumulation near sample boundaries.

The continuity equations were supplemented by appropriate boundary conditions, which should be fulfilled at the Co$_{20}$Fe$_{60}$B$_{20}|$Au interface and the outer boundaries of the Co$_{20}$Fe$_{60}$B$_{20}$/Au bilayer connected to a constant-current source via a gold nanoplate (Fig.~\ref{image1}). At the MgO$|$Co$_{20}$Fe$_{60}$B$_{20}$ interface, the projection $J_3$ of the charge current density \textbf{J} on the $x_3$ axis of our reference frame orthogonal to the interface was set equal to the density $J_0$ of the tunnel current. In addition, the vector \textbf{J} was taken to be parallel to the $x_3$ axis near the lateral faces of the Co$_{20}$Fe$_{60}$B$_{20}$/Au bilayer and at the contact with the Au nanoplate, where \textbf{J} satisfies the equality $\textbf{J} = (L_1/d)\textbf{J}_0$ involving the nanoplate thickness \textit{d} = 5 nm along the $x_1$ axis. At the Co$_{20}$Fe$_{60}$B$_{20}|$Au interface, we specified the spin-current density $\textbf{J}_s$ via the boundary condition $\textbf{e}_n \cdot \textbf{J}_s=\textbf{e}_n \cdot \textbf{J}_\textrm{\tiny{sp}} - (J_0 \hbar/2e)\textbf{p}_f$, where $\textbf{e}_n$ is the unit normal vector of the interface, and $\textbf{p}_f = (N_\uparrow - N_\downarrow)/(N_\uparrow + N_\downarrow) \textbf{m}$ is the spin polarization of the ferromagnetic layer defined by the densities of states of spin-up ($N_\uparrow$) and spin-down ($N_\downarrow$) electrons at the Fermi level ~\cite{Huang2008_242509}. Of course, the spin-current direction $\textbf{e}_s$ was taken to be parallel to the lateral faces of the Co$_{20}$Fe$_{60}$B$_{20}$/Au bilayer in the vicinity of these faces. 

The sought functions $\phi(\textbf{r},t)$ and $\textbf{P}(\textbf{r},t)$ were calculated numerically by solving the system of differential continuity equations with the aid of a finite-element  method. The calculations were performed in the quasistatic approximation ($\partial \rho / \partial t = \partial P / \partial t  = 0$), which is justified by the fact that the period $1/f \sim 1$ ns of the current-induced magnetization precession is much longer than the characteristic time of charge ($\sim$0.1 ps ~\cite{Harris1953_1114}) and spin ($\tau_\textrm{\tiny{sf}} < 100$ ps ~\cite{Elezzabi1996_3220}) equilibration. Since the size $L_2$ of the Co$_{20}$Fe$_{60}$B$_{20}$/Au bilayer along the $x_2$ axis is taken to be much smaller than the size $L_1$ along the $x_1$ one ($L_2/L_1 = 0.1$), variations of the potential $\phi$ and the spin polarization density \textbf{P} along the coordinate $x_2$ can be ignored. Therefore, we restricted our numerical calculations to the solution of a two-dimensional problem enabling us to determine the functions $\phi(x_1, x_3, t)$ and $\textbf{P}(x_1, x_3, t)$. In addition, only the component $J_{32}^\textrm{\tiny{sp}}$ of the pumped spin current was taken into account in the calculations, because it was found that the components $J_{31}^\textrm{\tiny{sp}}$ and $J_{33}^\textrm{\tiny{sp}}$ have a negligible effect on the sought output voltage $\Delta V(x_3, t)$ of the device. The thickness of Au overlayer along the $x_3$ axis was chosen to be much larger than the Au spin-diffusion length $\lambda_\textrm{\tiny{sd}} = \sqrt{D \tau_\textrm{\tiny{sf}}} = 35$ nm ~\cite{Mosendz2010_214403} and set equal to 400 nm.

In the numerical calculations, the conductivity $\sigma = e \mu n$ of Co$_{20}$Fe$_{60}$B$_{20}$ was taken to be $4.45 \times 10^5$ S m$^{-1}$ ~\cite{Fan2014_3042}, which yields the electron mobility $\mu = n^{-1} 2.8 \times 10^{26}$ m$^{-1}$ V$^{-1}$ s$^{-1}$. The anomalous Hall angle $\alpha_\textrm{\tiny{AH}} = 2 \alpha_\textrm{\tiny{SH}}$ and the spin polarization $p_f$ of Co$_{20}$Fe$_{60}$B$_{20}$ were assumed to be 0.02 ~\cite{Zhu2014_202404} and 0.53 ~\cite{Huang2008_242509}, respectively. For Au, the conductivity equals $4.5 \times 10^7$ S m$^{-1}$ ~\cite{Haynes2014}, which gives $\mu = 4.81 \times 10^{-3}$ m$^2$ V$^{-1}$ s$^{-1}$ and $D = 1.25 \times 10^{-4}$ m$^2$ s$^{-1}$. The spin-flip relaxation time $\tau_\textrm{\tiny{sf}}$ and the spin Hall angle of Au were taken to be 9.84 ps and 0.0035 ~\cite{Mosendz2010_214403}. It should be noted that the spin polarization density in the free Co$_{20}$Fe$_{60}$B$_{20}$ layer was assumed uniform to ensure consistency with the macrospin approximation used to describe the magnetization dynamics. 

\begin{figure}[h!]
\centering
\includegraphics[width=0.8\linewidth]{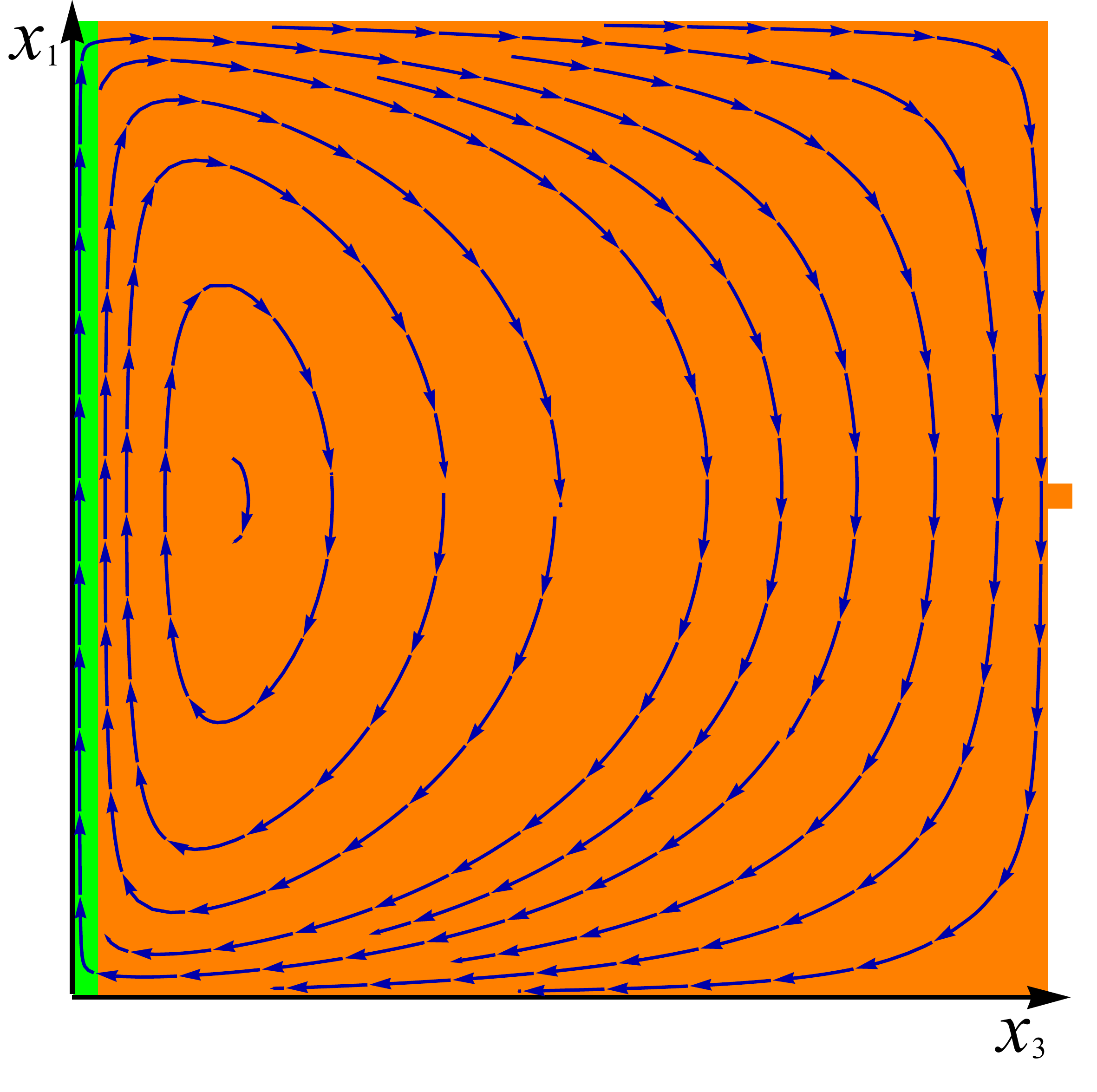}
\caption[]{ 
\justifying
Schematic diagram of the vortex-like contribution $\delta \textbf{J} = \textbf{J} - \textbf{J}_0$ to the charge current flowing in the Co$_{20}$Fe$_{60}$B$_{20}$/Au bilayer. The diagram shows the distribution $\delta \textbf{J} (x_1, x_3)$ at the current density $J_0 = J_\textrm{\tiny{min}}$ and the time moment, at which the free-layer magnetization has the direction cosines $m_1 = 0.881$, $m_2 = -0.473$, and $m_3 = 0$.}
\label{image6}
\end{figure}

\begin{figure}[h!]
\centering
\includegraphics[width=0.93\linewidth]{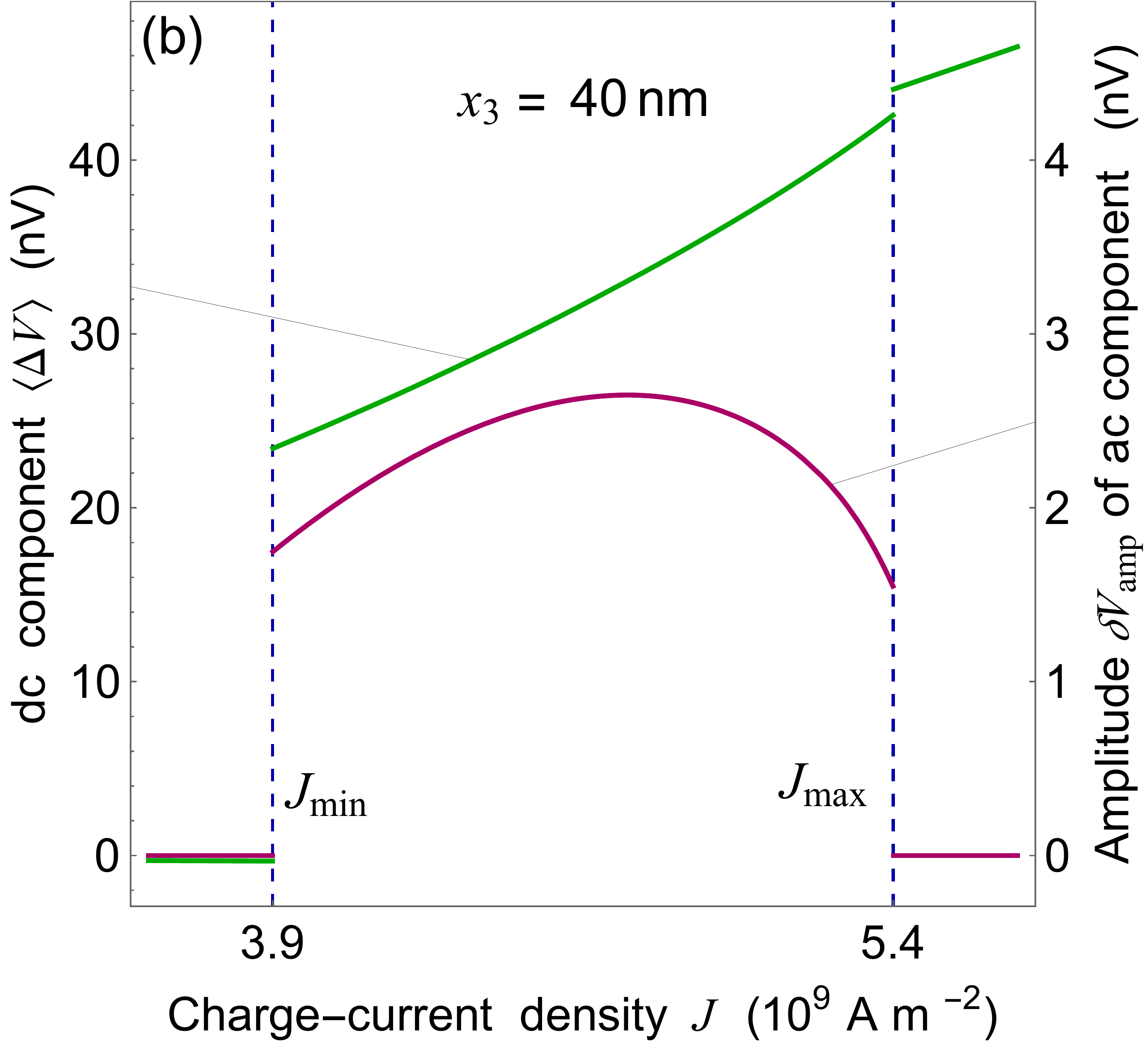}
\caption[]{ 
\justifying
Transverse voltage signal $\Delta V$ generated by the Co$_{20}$Fe$_{60}$B$_{20}$/Au bilayer. (a) Variations of the dc component $\left\langle \Delta V \right\rangle$ of this signal with the distance $x_3$ from MgO$|$Co$_{20}$Fe$_{60}$B$_{20}$ interface calculated at different densities of the applied charge current for the steady-state magnetization precession. (b) Dependences of the dc component $\left\langle \Delta V \right\rangle$ and the amplitude $\delta V_\textrm{\tiny{amp}}$ of the ac component at the point $x_3 = 40$ nm inside the Au layer on the charge-current density \textit{J}.}
\label{image7}
\end{figure}

Using the obtained functions $\phi(x_1,x_3,t)$ and $\textbf{P}(x_1,x_3,t)$, we calculated spatial distributions of the charge-current density $\textbf{J}(x_1,x_3,t)$ in the Co$_{20}$Fe$_{60}$B$_{20}$/Au bilayer and the electrical potential difference $\Delta V(x_3, t)$ between its lateral sides. Interestingly, the charge-current distribution at any fixed moment \textit{t} can be represented as a sum of the applied uniform current $\textbf{J}_0$ and a vortex-like contribution $\delta \textbf{J}(x_1,x_3,t)$ illustrated by Fig.~\ref{image6}. The transverse voltage signal $\Delta V(x_3, t)$ generated by the device decreases with increasing distance $x_3$ from the MgO$|$Co$_{20}$Fe$_{60}$B$_{20}$ interface, falling rapidly within the Co$_{20}$Fe$_{60}$B$_{20}$ layer [Fig.~\ref{image7}(a)].

\begin{figure}[h!]
\flushleft
\includegraphics[width=0.88\linewidth]{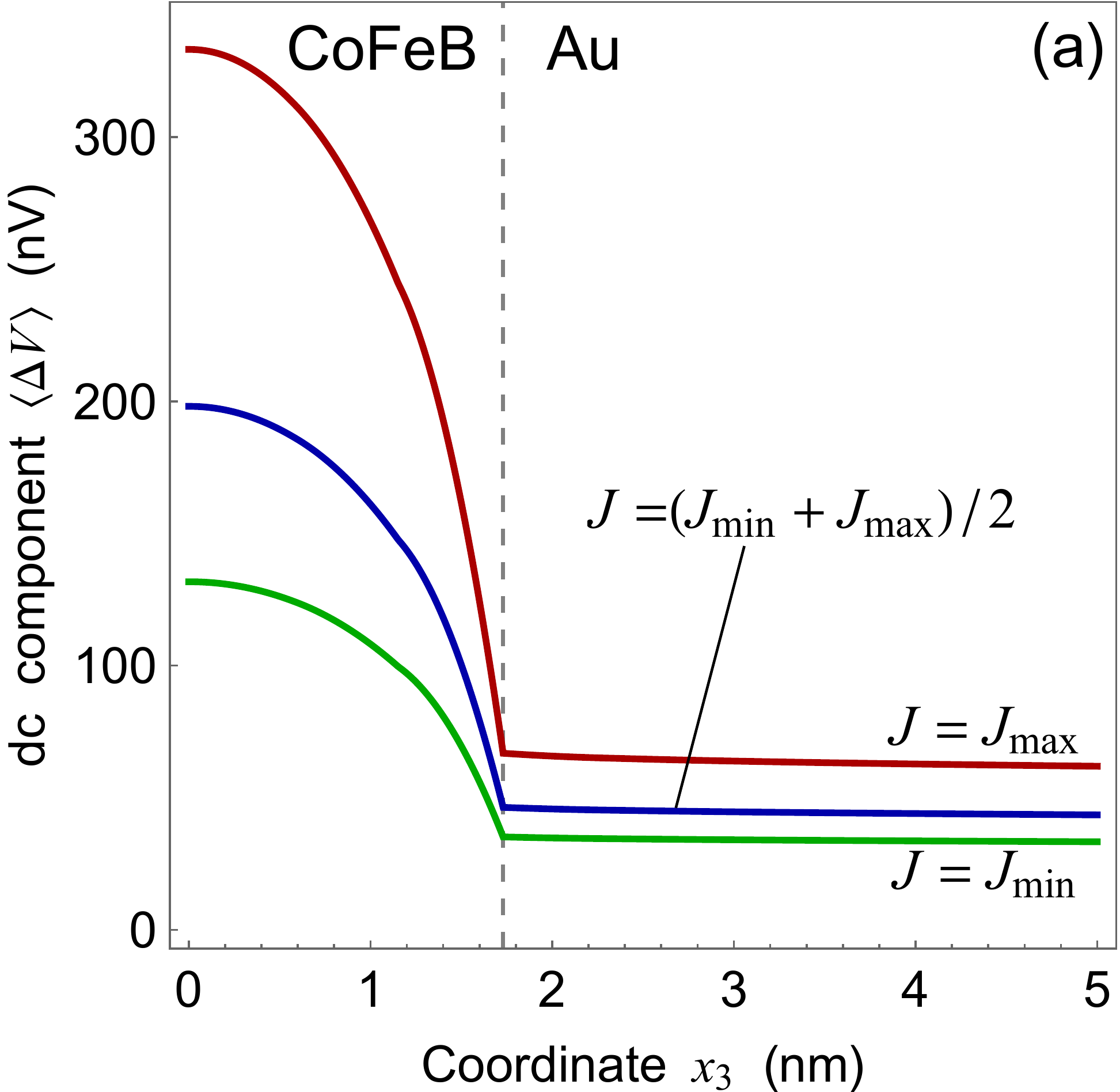}
\label{ris:image1}
\end{figure}

\begin{figure*}[t]
\centering
\begin{minipage}[h]{0.5\linewidth}
\includegraphics[width=0.9\linewidth]{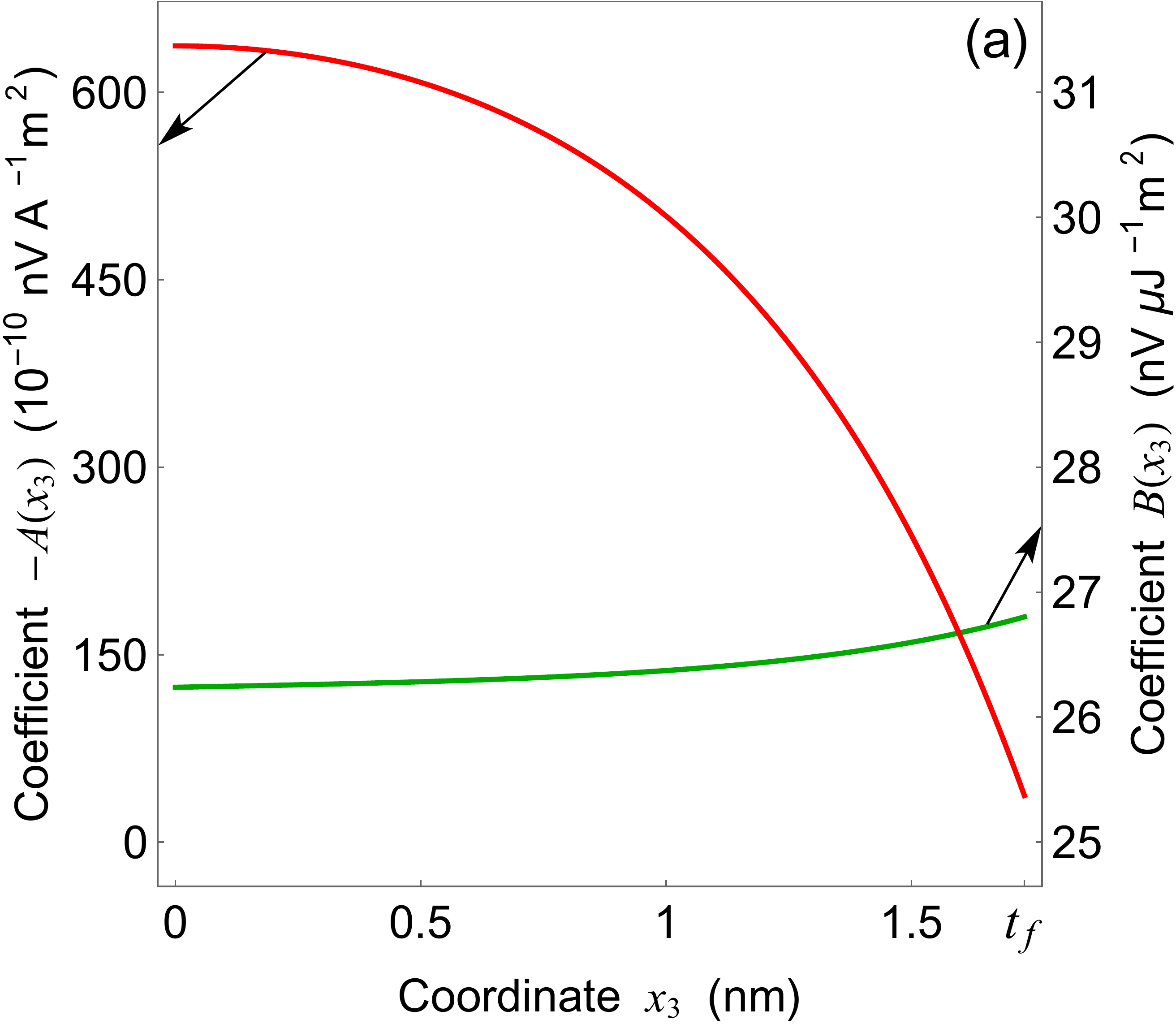}
\end{minipage}
\hfill
\begin{minipage}[h]{0.49\linewidth}
\includegraphics[width=0.9\linewidth]{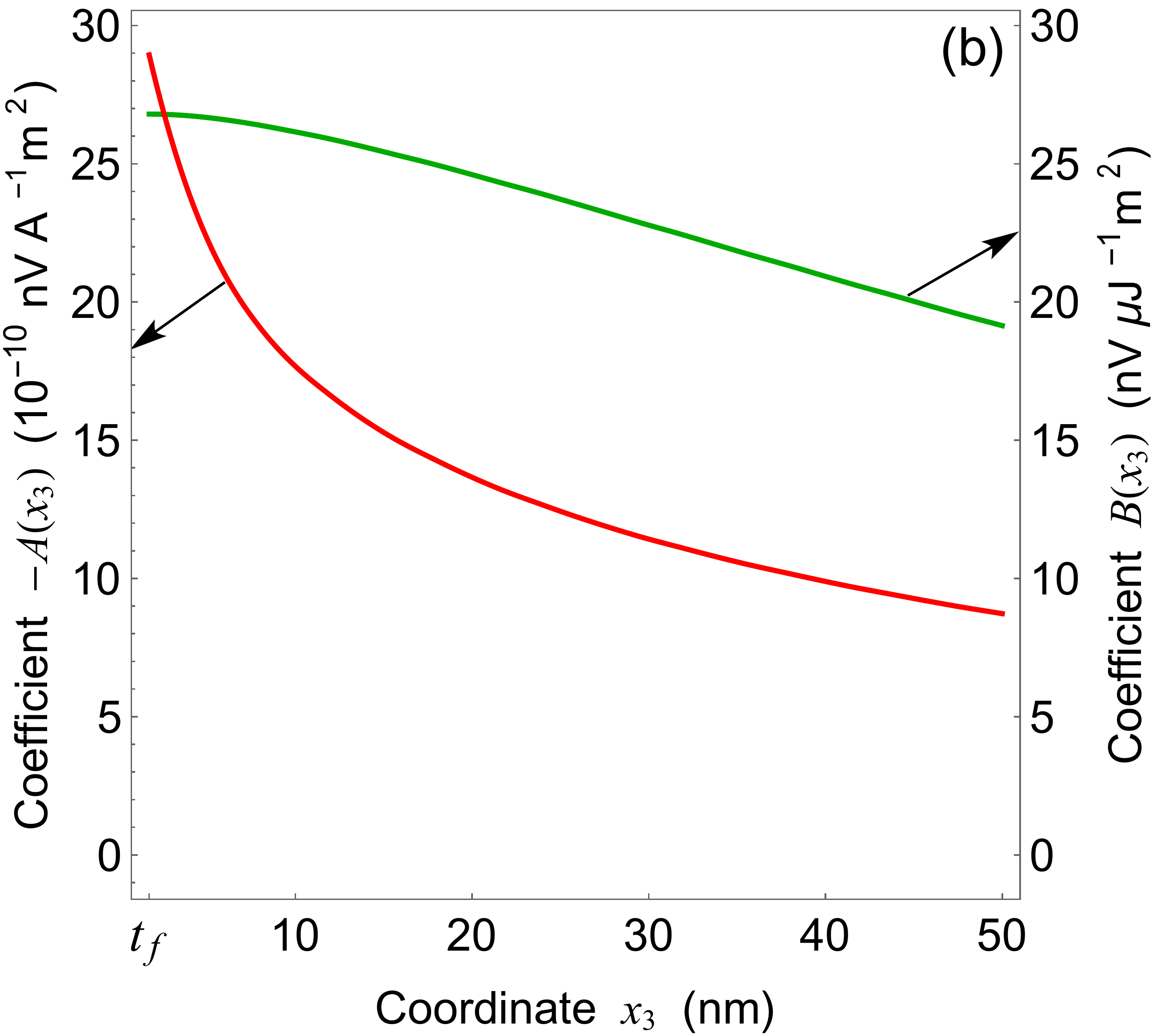}
\end{minipage}
\caption[]{ 
\justifying
Dependences of the coefficients \textit{A} and \textit{B} involved in Eq. \hyperref[eq6]{(6)} on the distance $x_3$ from the MgO$|$Co$_{20}$Fe$_{60}$B$_{20}$ interface within the free Co$_{20}$Fe$_{60}$B$_{20}$ layer (a) and in the Au overlayer (b).}
\label{image8}
\end{figure*}

\begin{figure}[h!]
\centering
\includegraphics[width=0.9\linewidth]{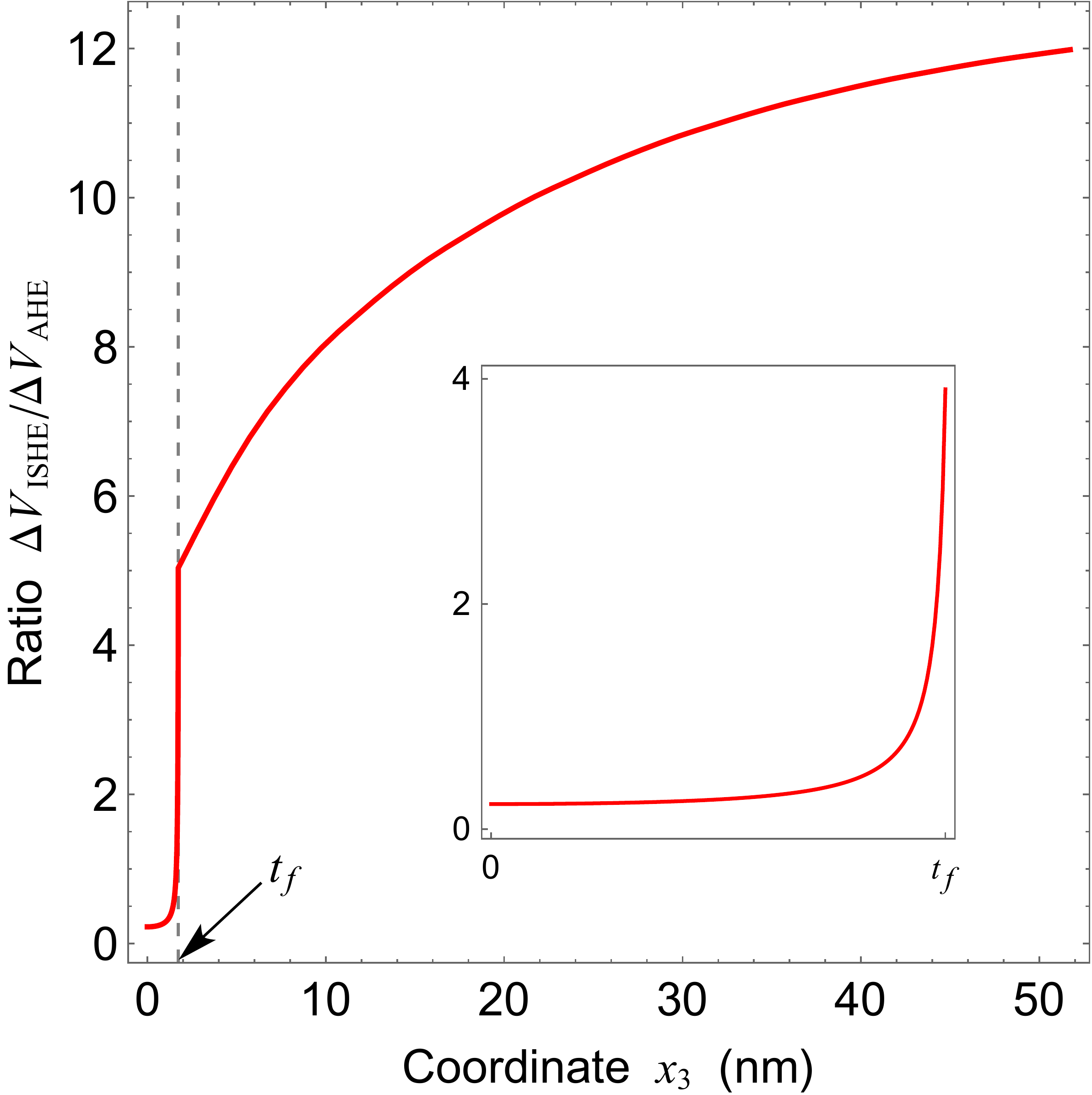}
\caption[]{ 
\justifying
Ratio $\Delta V_\textrm{\tiny{ISHE}} / \Delta V_\textrm{\tiny{AHE}}$ plotted as a function of the distance $x_3$ from the MgO$|$Co$_{20}$Fe$_{60}$B$_{20}$ interface. The presented curve corresponds to the charge-current density $J = 4.72 \times 10^9$ A m$^{-2}$, at which the ac contribution to the total spin injection into the Au layer reaches maximum. The inset shows the variation of $\Delta V_\textrm{\tiny{ISHE}} / \Delta V_\textrm{\tiny{AHE}}$ inside the Co$_{20}$Fe$_{60}$B$_{20}$ layer of thickness $t_f = 1.73$ nm.}
\label{image9}
\end{figure}

\newpage
\noindent
Remarkably, the analysis of the numerical results obtained for the transverse voltage reveals that $\Delta V(x_3, t)$ can be fitted with a high accuracy by the analytical formula

\begin{equation}
\label{eq6}
\begin{gathered}
\Delta V(x_3, t)=A(x_3)m_2(t)J_0+B(x_3)J_{32}^s(x_3=t_f,t),
\end{gathered}
\end{equation}

\noindent
where the first term represents the contribution $\Delta V_\textrm{\tiny{AHE}}$ of the anomalous Hall effect, while the second term describes the contribution $\Delta V_\textrm{\tiny{ISHE}}$ resulting from the inverse spin Hall effect. Since the coefficients $A(x_3)$ and $B(x_3)$ involved in Eq. \hyperref[eq6]{(6)} have very different dependences on the distance $x_3$ (see Fig.~\ref{image8}), the ratio $\Delta V_\textrm{\tiny{ISHE}} / \Delta V_\textrm{\tiny{AHE}}$ changes strongly across the Co$_{20}$Fe$_{60}$B$_{20}|$Au interface. Figure~\ref{image9} demonstrates that this ratio is mostly very small inside the Co$_{20}$Fe$_{60}$B$_{20}$ layer, but rises steeply near the Co$_{20}$Fe$_{60}$B$_{20}|$Au interface and exceeds 5 in the Au layer. Hence, measurements of the average voltage signal created by the Co$_{20}$Fe$_{60}$B$_{20}$ layer provide information on the anomalous Hall effect, whereas the potential difference $\Delta V(x_3, t)$ between the faces of the Au layer measured at $x_3 > 25$ nm characterizes the inverse spin Hall effect.

\begin{figure*}[t]
\centering
\begin{minipage}[h]{0.5\linewidth}
\includegraphics[width=0.9\linewidth]{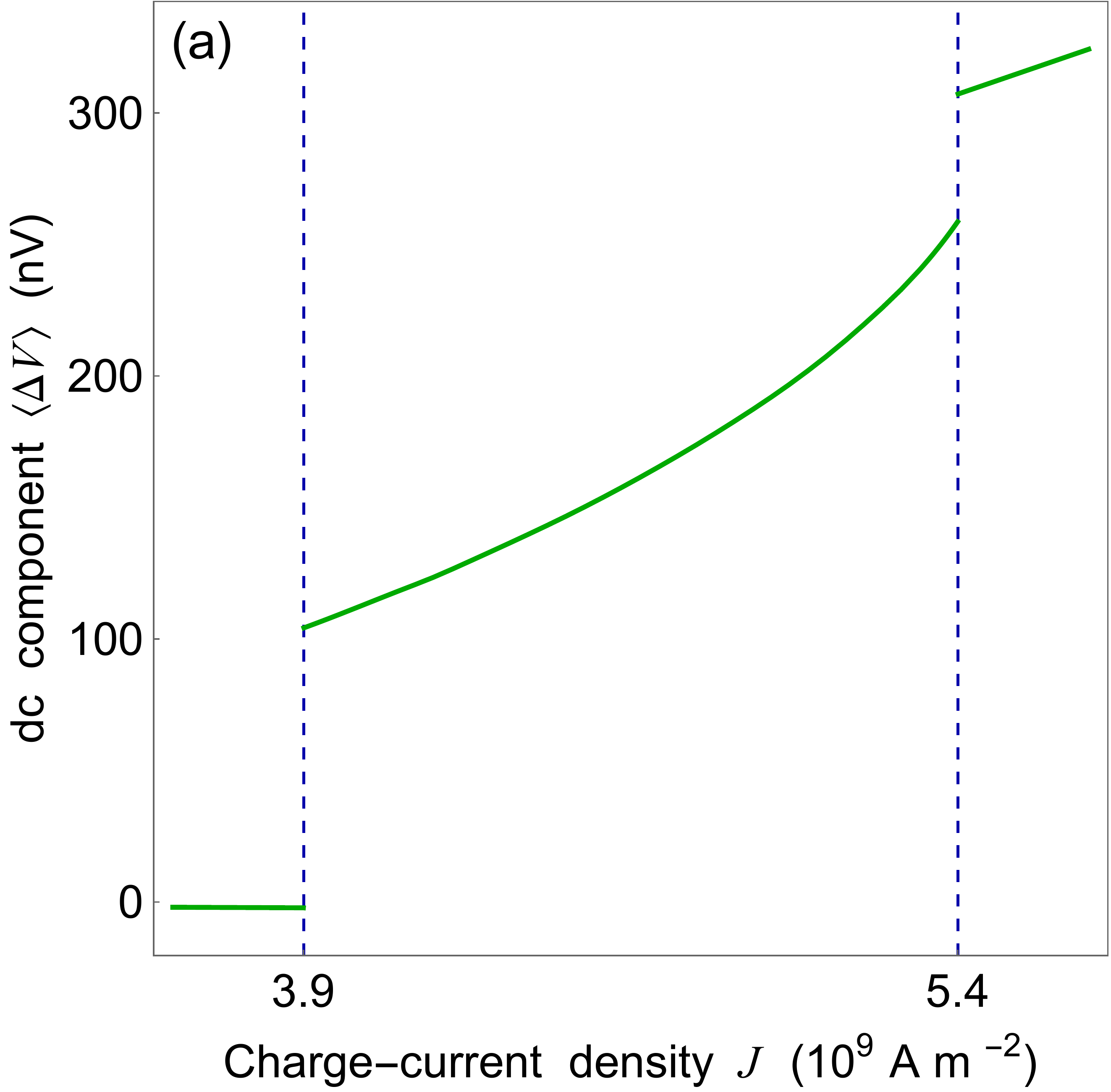}
\end{minipage}
\hfill
\begin{minipage}[h]{0.49\linewidth}
\includegraphics[width=0.9\linewidth]{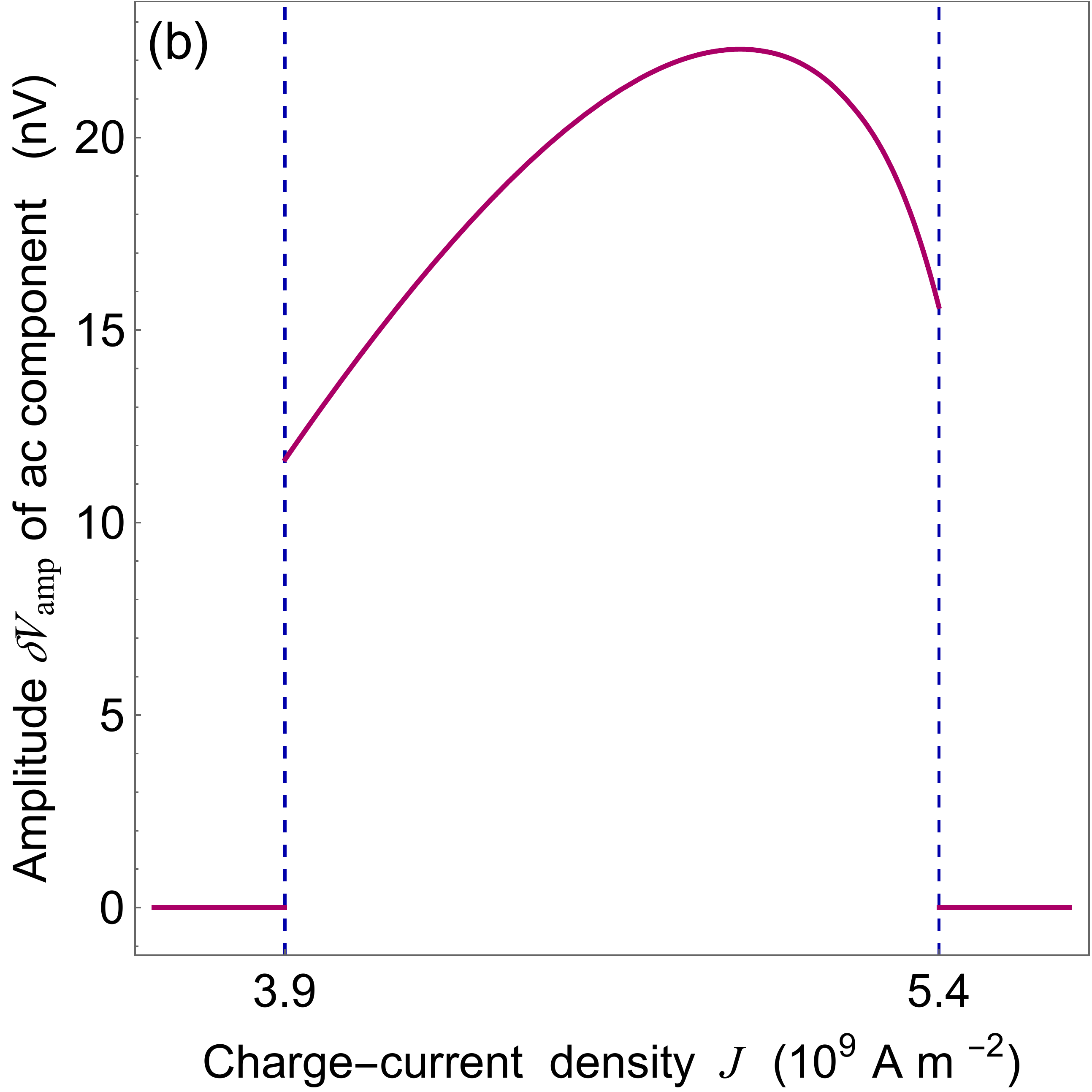}
\end{minipage}
\caption[]{ 
\justifying
Transverse voltage signal $\Delta V$ averaged over the Co$_{20}$Fe$_{60}$B$_{20}$ layer plotted as a function of the charge-current density \textit{J}. Panels (a) and (b) show the dc component $\left\langle \Delta V \right\rangle$ and the amplitude $\delta V_\textrm{\tiny{amp}}$ of the ac component of this signal, respectively.}
\label{image10}
\end{figure*}

Figure~\ref{image10} shows how the dc and ac components of the transverse signal $\Delta V$ averaged over the thickness $t_f$ of the Co$_{20}$Fe$_{60}$B$_{20}$ layer vary with the charge-current density \textit{J}. It can be seen that the curves are similar to the dependences $J_{32}^\textrm{\tiny{sc}}(J)$ presented in Fig.~\ref{image5}, which describe the spin injection into Au caused by the spin-polarized charge current. In contrast, Fig.~\ref{image7}(b) demonstrates the dc component $\left\langle \Delta V \right\rangle(J)$ and the amplitude $\delta V_\textrm{\tiny{amp}}(J)$ of the GHz-frequency ac component calculated at $x_3 = 40$ nm. Importantly, both the dc and microwave signals appear to be large enough for the experimental detection within the precession window. Moreover, the dependences $\left\langle \Delta V \right\rangle(J)$ and $\delta V_\textrm{\tiny{amp}}(J)$ repeat the graphs shown in Fig.~\ref{image5} for the total spin-current density $J_{32}^s$ generated at the Au$|$Co$_{20}$Fe$_{60}$B$_{20}$ interface, differing by a constant factor of 20.61 nV $\mu$J$^{-1}$ m$^2$ only. Hence the measurements of $\Delta V$ by nanocontacts placed at distances $\delta x_3 \sim \lambda_\textrm{\tiny{sd}}$ from the boundary of ferromagnetic layer provide information on the spin injection into the normal metal.

\section{\label{sec:four}SUMMARY}
In summary, we presented a comprehensive theoretical study of the spin dynamics and charge transport in the Co$_{20}$Fe$_{60}$B$_{20}$/MgO/Co$_{20}$Fe$_{60}$B$_{20}$/Au tunneling heterostructure connected to a constant-current source. The performed numerical calculations enabled us to find the range of current densities, within which electrically driven magnetization precession appears in the free Co$_{20}$Fe$_{60}$B$_{20}$ layer, and to determine the precession frequencies and trajectories. Remarkably, a novel, dynamic SRT has been predicted, which is caused by the joint impact of STT and VCMA and has the form of magnetization
reorientation between initial static direction and final dynamic
precessional state. The results obtained for the magnetization dynamics were then used to evaluate the dc and ac components of the spin current generated in the Au overlayer owing to the precession-driven spin pumping and the oscillating spin polarization of the charge current. Taking into account the inverse spin Hall effects and anomalous Hall effects, we finally calculated the charge flow and electric-potential distribution in the Co$_{20}$Fe$_{60}$B$_{20}$/Au bilayer via the numerical solution of coupled drift-diffusion equations for charge and spin currents. It is shown that the potential difference between the lateral faces of the Au layer is large enough for experimental detection, which demonstrates significant efficiency of the described spin injector.

\section*{\label{sec:four}\\ACKNOWLEDGMENTS}
The work was supported by the Foundation for the Advancement of Theoretical Physics and Mathematics "BASIS".

\end{document}